\documentclass[journal, final]{IEEEtran}

\usepackage{textpos}
\newcommand\copyrighttext{
\small This article has been accepted for publication in IEEE Transactions on Wireless Communications.  

}
\newcommand\copyrightnotice{%
	\begin{tikzpicture}[remember picture,overlay]
		\node[anchor=north,yshift=-15pt] at (current page.north)  {\copyrighttext}
		;
	\end{tikzpicture}%
}

\usepackage{setspace}
\usepackage{amsmath}
\usepackage{bm}
\usepackage{mdframed}
\usepackage{amsthm}

\usepackage[utf8]{inputenc}
\usepackage{graphicx}
\usepackage[colorinlistoftodos]{todonotes}
\usepackage{amssymb}
\usepackage{longtable}
\usepackage[colorlinks=true, allcolors=blue]{hyperref}
\usepackage[noadjust]{cite}
\allowdisplaybreaks
\usepackage{lineno}
\usepackage{lipsum}
\usepackage{multicol}
\usepackage{subcaption}
\usepackage{enumitem}
\newtheorem{theorem}{Theorem}

\newtheorem{assumption}{Assumption}
\usepackage{multirow}
\usepackage{array}

\usepackage{xcolor}

\usepackage[vlined,linesnumbered,ruled,resetcount]{algorithm2e}

\newcommand{\secrev}[1]{{\color{black}#1}}
\newcommand{\thirdrev}[1]{{\color{black}#1}}

\SetKwInput{KwInput}{Input}                
\SetKwInput{KwOutput}{Output}              
\SetKwInput{KwInitialize}{Initialize}

\usepackage{booktabs}
\usepackage{caption}
\usepackage{float}
\usepackage{capt-of}
\usepackage{arydshln}
\usepackage{soul,xcolor}
\usepackage{tabularx}

\usepackage{arydshln}
\title{Constrained Deep Reinforcement Based Functional Split Optimization in Virtualized RANs}

\author{\IEEEauthorblockN{Fahri Wisnu Murti, Samad Ali, and Matti Latva-aho}\\
\IEEEauthorblockA{
	Centre for Wireless Communications, University of Oulu, Finland}
\thanks{A preliminary version of this work appears in IEEE ICC 2021 Workshop \cite{murti_iccw_cdrs_vran}. This research has been supported by the Academy of Finland, 6G Flagship program under Grant 346208.}%
}

\IEEEoverridecommandlockouts
\begin{document}
	

\maketitle

\begin{abstract}
\thirdrev{In virtualized radio access network (vRAN), the base station (BS) functions are decomposed into virtualized components that can be hosted at the centralized unit or distributed units through functional splits. Such flexibility has many benefits; however, it also requires solving the problem of finding the optimal splits of functions of the BSs in such a way that minimizes the total network cost. The underlying vRAN system is complex and precise modelling of it is not trivial. Formulating the functional split problem to minimize the cost results in a combinatorial problem that is provably NP-hard, and solving it is computationally expensive. In this paper, a constrained deep reinforcement learning (RL) approach is proposed to solve the problem with minimal assumptions about the underlying system. Since in deep RL, the action selection is the outcome of inference of a neural network, it can be done in real-time while training to update the neural networks can be done in the background. However, since the problem is combinatorial, even for a small number of functions, the action space of the RL problem becomes large. Therefore, to deal with such a large action space, a chain rule-based stochastic policy is exploited in which a long short-term memory (LSTM) network-based sequence-to-sequence model is applied to estimate the policy that is selecting the functional split actions. However, the utilized policy is still limited to an unconstrained problem, and each split decision is bounded by vRAN’s constraint requirements. Hence, a constrained policy gradient method is leveraged to train and guide the policy toward constraint satisfaction. Further, a search strategy by greedy decoding or temperature sampling is utilized to improve the optimality performance at the test time. Simulations are performed to evaluate the performance of the proposed solution using synthetic and real network datasets. Our numerical results show that the proposed RL solution architecture successfully learns to make optimal functional split decisions with the accuracy of the solution is up to 0.05\% of the optimality gap. Moreover, our solution can achieve considerable cost savings compared to C-RAN or D-RAN systems and a faster computational time than the optimal baseline.}

\end{abstract}

\IEEEpeerreviewmaketitle

\copyrightnotice
\vspace{-2mm}

\section{Introduction}

The increase in mobile data trafﬁc of emerging applications with diverse requirements has driven the efforts to re-design the radio access networks (RANs). Cloud/Centralized-RAN (C-RAN) has become a favourable solution to enable the low-cost deployment and high-performance systems by pooling the baseband functions of the base station (BS) to a central server which is also known as Cloud/Central unit (CU). \thirdrev{This idea is motivated by the growth of network densification, enabled by the concept of \textit{softwarization}, to offer cost-efﬁcient solutions and high-performance network operations through centralized control \cite{cost_vm}.} However, a fully centralized RAN \thirdrev{is typically not implementable for many reasons \cite{vran_murti2}}. For instance, it requires a low-latency and high-capacity fronthaul, \thirdrev{which is frequently absent in current RANs and prohibitively expensive to develop from scratch. This challenge motivates the transition from} rigid C-RAN designs to flexible architectures, where only a subset of BS functions is centralized at the CU, and the other functions are hosted at the distributed units (DUs) \thirdrev{and radio units (RUs)}\footnote{\thirdrev{RUs are the radio hardware units to run the RF functions.}}. Further, the term virtualized RAN (vRAN) \thirdrev{is coined to describe these architectures \cite{openvran_nec}.}

\thirdrev{In vRANs, the BS functions (except RF functions) can be decomposed into virtualized components and executed on commodity hardware across a geo-distributed edge cloud system\footnote{Each CU and DU \thirdrev{is} to run as virtualized software, e.g., virtualized CU (vCU) and virtualized DU (vDU). A DU is typically executed at the far-edge server (co-located or close to the RU), while a CU is \thirdrev{at the edge server (a more centralized server)}.} \cite{nokia_cran_edge}.
 Then, the operators can uniquely select the \textit{functional splits} suited to their needs by deciding which functions will be centralized at the CU and which will be kept at the DUs. This paradigm brings flexibility to the RAN operations, potentially offers a cost-saving, and accommodates diverse use cases and applications in 5G+ systems \cite{function_split_survey, nokia_anyhaul}.} 
%
%
%
%
%
%
%
However, selecting the functional splits of all the BSs is challenging. Each split \secrev{has a different delay requirement, initiates a different computing load to the CU and DUs, and induces a different data flow}. The initial design of vRAN fronthaul using point-to-point Common Public Radio Interface (CPRI) \thirdrev{is also suggested to be updated with the new} Crosshaul/xHaul \thirdrev{architecture} based on an open interface and packet-switch (shared) network \thirdrev{such as the enhanced CPRI} \cite{function_split_survey, nokia_anyhaul, xhaul}.
\thirdrev{As a result, in addition to sharing the same computing nodes, each BS has to share the same network links, \thirdrev{which leads to complex interdependence between split decisions.} 
Therefore, it is important to optimize the splits carefully to ensure that the deployment is beneficial; otherwise, it can lead to high operating costs and performance degradation. } 

\secrev{On the other hand, optimizing the functional splits produces a high degree of complexity. In addition to \thirdrev{the mentioned challenges, unlike legacy RANs, the behaviour of vRAN system performance such as computing utilization \cite{vranai_journal} and power consumption \cite{jose_bayesian} is highly non-trivial. This non-triviality is also reinforced by vRAN deployment over the same platform with other workloads such as video analytics \cite{concordia_vran}.} As a result, it is \thirdrev{complex and difficult }to model the underlying system precisely. Meanwhile, traditional mathematical optimization approaches rely on complete knowledge of the system behaviour to define the models and solve the problems; \thirdrev{and this }can be unfeasible in practice. These challenges motivate us to shift to machine learning (ML)-based approaches, which can be best to address our vRAN splitting problem amidst minimal assumptions about the underlying system.} 

\vspace{-1mm}
\subsection{Related Works}

\secrev{\textbf{Optimization-based approaches.}} 3GPP \cite{split_3gpp,split_3gpp_rel16} and a seminal white paper \cite{smallcell} have defined the detail vRAN split specifications. Although the authors in  \cite{function_split_survey} have discussed the gains and requirements for each split, there are still few works on the optimization issues. Energy consumption for various splits has been evaluated in \cite{apt-ran}.
The authors in \cite{wizhaul_andres} have studied optimizing the function centralization of vRANs over xHaul. Follow-up works, \cite{fluidran_andres} and \cite{vranmec_andres} offered an optimal solution of minimizing total cost for integration vRANs with Mobile Edge Computing (MEC). Then, the work in \cite{vran_murti1} has proposed an optimized multi-cloud vRAN framework with balancing its centralization \cite{vran_murti2}. The authors in \cite{placeran} have proposed the PlaceRAN framework to minimize the computing resources while maximizing the radio function aggregations \secrev{using the IBM CPLEX solver}. These mentioned works \cite{apt-ran,wizhaul_andres,fluidran_andres,vranmec_andres,vran_murti1,vran_murti2,placeran} have addressed various optimization problems in vRANs. \secrev{However, their frameworks require assuming complete models of the underlying system to deﬁne their problem structures and solve the problems. We argue that such strong assumptions can be inaccurate as the underlying vRAN system is complex and difficult to model precisely.} Additionally, those \secrev{frameworks} need heavy mathematical solutions \thirdrev{with} exponential complexity and slow execution time, \thirdrev{which typically are unsuitable} for large networks and online \thirdrev{execution}. The above problems are \thirdrev{also} often complex combinatorial and difficult to solve \secrev{optimally}. \secrev{Therefore}, we opt out to use optimization-based approaches to \secrev{formulate and solve our vRAN splitting problem}.



\secrev{\textbf{ML-based approaches.} ML techniques recently have been spurred to address complex optimization and control problems in wireless networks \cite{ali20206g, ali2020ml}. The authors in \cite{vranai_journal} have proposed a learning framework that successfully solves a contextual bandit problem of dynamic computing and radio resource controls in vRANs using a deep reinforcement learning (RL) paradigm. Further, they leveraged Bayesian learning for energy efficient-based resource orchestrator in \cite{jose_bayesian}. ML-based predictor also has been developed in \cite{concordia_vran} that learns to share the CPU resources between a vRAN workflow and other workflows in the same server. Although the authors in \cite{vranai_journal,jose_bayesian,concordia_vran} have shown the non-triviality of vRAN performance  and the importance of learning based-framework to manage the vRAN resources, they still did not discuss how to design a framework that learns to optimize the functional splits. 
}
Recent work in \cite{vran_dl} has studied user-centric slicing and split optimization problems using a deep learning \secrev{method}. The authors modelled their problem as supervised learning, which relies on high-quality labelled datasets (e.g., optimal labels) to assess the quality of the decisions. 
Once trained, the model can be used quickly in an online manner, offering a real-time solution \secrev{for assigning the split for each user slice}. However, in vRANs, obtaining such high-quality labels is expensive. \thirdrev{To construct the labelled datasets, we still need} complete knowledge of the system performance to model the problem mathematically \thirdrev{and} solve multiple instances of the problem. 
\secrev{The work in \cite{green_oran} has addressed the functional split problem for green Open RANs using Q-learning and SARSA; however, they assumed each DU/CU as an independent agent that focuses on its own utility. We argue that every split decision in vRANs is interdependent as the BSs share the same network links and computing nodes with limited capacity. Besides, \cite{vran_dl} focused on the split assignment for the users, and \cite{green_oran} studied the effectiveness of energy sources, but we aim for a different goal.}

\vspace{-1mm}
\subsection{Methodology \& Contributions}

Our goal is to develop a zero-touch optimization framework that optimizes the functional \secrev{splits} \thirdrev{of} \secrev{the BSs to minimize the total network cost} in the vRAN system. First, we formulate and present the functional split problem mathematically to provide a better understanding of its objective and constraints. Our formulation yields a combinatorial and NP-hard problem. Therefore, it is computationally expensive to solve optimally, especially for large-scale networks and real-time execution. Moreover, solving such a problem often relies on \secrev{the assumptions of the underlying system to define and model the problem structure (e.g., mathematical optimization). However, in practice, the behaviour of vRAN performance and resources is highly non-trivial, which is complex and can be unfeasible to model the system precisely. }   
 
Motivated by the above challenges, we formulate the functional split problem with {constrained neural combinatorial RL}. We use neural networks \secrev{to approximate the policy that} maps the state observations to the actions. Then, the idea is to estimate the neural network model's parameters iteratively by taking instances from the problem spaces using a constrained deep RL paradigm. For every interaction with the environment (vRAN system), we expect to receive a reward (the induced total network cost) and penalty (constraint violation) as feedback signals and the output returned by the neural network to learn and improve the model. \thirdrev{This paradigm} considers the vRAN system as a black-box environment, \thirdrev{making} minimal assumptions about the underlying system. It also does not need the optimal labelled datasets, which are highly expensive to obtain in vRANs.

\thirdrev{Further, we propose a novel constrained deep reinforcement-based functional split optimization framework (CDRS) to solve the problem. Due to the combinatorial nature of the problem finding the optimal splits, the action space of our RL problem becomes enormously large. Therefore, we develop CDRS using a chain rule-based stochastic policy \cite{neural_bello} in which policy network architecture using a long short-term memory (LSTM) network-based sequence-to-sequence model is applied to estimate the policy \cite{seq2seq,attention_bahdanau}. However, this policy is still limited to an unconstrained problem, which is not directly applicable to our vRAN splitting problem. Therefore, we leverage a constrained policy gradient method to train and guide the policy toward constraint satisfaction. Then, CDRS can be tailored into {CDRS-Fixed} and {CDRS-Ada}. CDRS-Fixed uses a fixed penalty coefficient \cite{solozabal_constrained, vnf_drl_solozabal} while {CDRS-Ada} updates the penalty coefficient adaptively \cite{reward_constraint,pdo_risk}. A self-competing baseline is also utilized with an auxiliary network to improve the policy further. Once the model is trained, finding the solution for the problem is computationally efficient as it only requires a forward pass through the trained neural network. Therefore, we provide a search strategy to improve the optimality performance at the test time. Following the search strategy, CDRS can be further tailored into {CDRS-Fixed-G}, {CDRS-Ada-G}, {CDRS-Fixed-T} and  {CDRS-Ada-T}. CDRS-Fixed-G and CDRS-Ada-G utilize greedy decoding while CDRS-Fixed-T and CDRS-Ada-T use temperature sampling.}

\thirdrev{CDRS is evaluated} in terms of training behaviour, optimality performance, the impact of altering the traffic load and routing cost and the computational time. The evaluations are performed using a synthetic network generated by the Waxman algorithm \cite{waxman} and a real network dataset \cite{sndlib}. The used system parameters are from a measurement-based 3GPP-compliant system model. To assess the effectiveness of our approach, we compare it to the optimal value obtained from a Phyton-MIP solver\footnote{We use a solution obtained from a mixed-integer programming solver (https://www.python-mip.com/) as an optimal baseline comparison. It offers an exact solution through a well-known method, Branch-\&-Cut algorithm. }. Following our evaluations, CDRS successfully learns the optimal \secrev{functional splits} and solves the problem with $0.05\%$ of the optimality gap\footnote{We use the term \emph{optimality gap} to define our solution's error compared to the optimal value obtained from the MIP solver.} (e.g., CDRS-Fixed-T).  Our results also show that CDRS is the most cost-effective compared to two extreme cases: fully C-RAN and D-RAN. All of our CDRS settings are \thirdrev{also} faster than the MIP solver, where CDRS-Ada-G can attain as high as 22.82 times faster. Our contributions can be summarized:
\begin{itemize}
	\item  We formulate the vRAN split problem to constrained neural combinatorial RL, which takes minimal assumptions about the underlying system and does not require the optimal labelled datasets to solve the problem. We also consider the interdependence between split decisions capturing the network links and computing resources sharing among the BSs.
	
	
	\item We propose CDRS as \thirdrev{a novel solution framework}. CDRS adopts a chain rule-based stochastic policy \thirdrev{to} deal with the interdependence between split decisions and the combinatorially large discrete action space of the problem in which an LSTM networks\thirdrev{-based sequence-to-sequence model} is applied to estimate the policy. \thirdrev{We utilize} a constrained policy gradient method with a self-competing baseline \thirdrev{to train and guide the policy toward constraint satisfaction.} Following the penalty coefﬁcient and search strategy settings, \thirdrev{CDRS can be tailored} into CDRS-Fixed-G, CDRS-Fixed-T, CDRS-Ada-G and CDRS-Ada-T.

	\item We conduct a series of evaluations using synthetic and real network datasets. We investigate the training behaviour, the accuracy of the solution, the impact of routing cost and traffic load and the computational time.  
\end{itemize}


To the best of our knowledge, this work is ﬁrst to optimize the functional splits of the BSs to minimize the total network cost in the vRAN system using a constrained deep RL paradigm, which takes minimal assumptions about the underlying system and adopts a chain rule-based stochastic policy to deal with the large action space and interdependence between decisions.




The rest of this paper is organized as follows. The background and system model of vRAN are presented in Section \ref{sec:model}. The functional split problem is formalized mathematically in Section \ref{sec:problem}. Our proposed framework is described in Section \ref{sec:solution}. Our simulation and experiment results are discussed in Section \ref{sec:results}. Finally, our work is concluded in Section \ref{sec:conclusion}.


\vspace{-1mm}
\section{System Model} \label{sec:model}

\begin{figure}[t!]
	\centering
	\includegraphics[width=0.45 \textwidth]{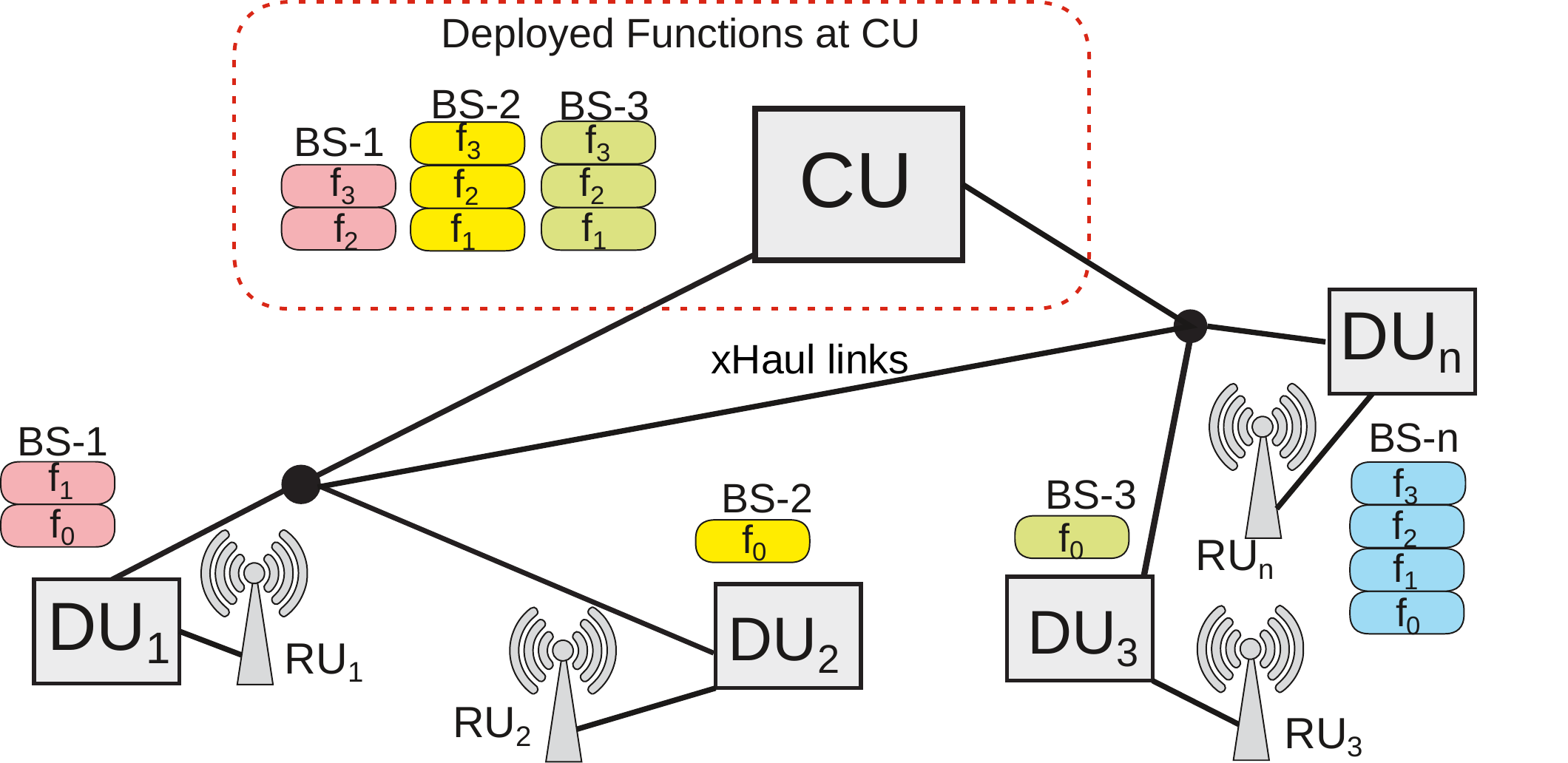}   
	\caption{\small vRAN over integrated fronthaul/midhaul (xHaul). It has many degrees of design freedom by possibly hosting BS functions at the CU or DUs.}
	\label{fig:vran}
	\vspace{-3mm}
\end{figure}

\textbf{Background.} In C-RAN, all BS functions are centralized at the Base Band Unit (BBU) except RF layers at the RU.  \thirdrev{In vRANs}, the BBU is decoupled into the CU and DU \cite{split_3gpp_rel16}. Hence, functions of a BS can be deployed at the CU, DU and RU. Fig \ref{fig:vran} illustrates that a CU is typically executed at a bigger and more centralized \thirdrev{server} (e.g., edge server), while a DU is at a smaller server (e.g., far-edge server) and located near (or co-located) with an RU.

Our model refers to the standardization of 3GPP \cite{split_3gpp,split_3gpp_rel16} and seminal white paper \cite{smallcell}, where each split has a different performance gain \cite{vran_murti2,function_split_survey}. \thirdrev{Although 3GPP has defined eight options for the splits, several are still hardly implemented. Therefore, we consider four splits that have been experimentally validated in a prototype \cite{costdu_nikaein,adaptive_alba}. }
 \textbf{Split 0}: All functions are at the DU, except the RF layers at the RU. It is a typical D-RAN setup. \textbf{Split 1} (PDCP-RLC): RRC, PDCP, and upper layers are hosted at the CU, while RLC, MAC, and PHY are at the DU. \secrev{This split enables a separate user plane \thirdrev{and control plane with} centralized RRC.} \textbf{Split 2} (MAC-PHY): MAC and upper layers are at the CU, while PHY is at the DU. It allows improvement for CoMP by centralized HARQ. \textbf{Split 3} (PHY-RF): All functions are at the CU, except RF layers. It is a fully centralized version of vRANs. It gains power-saving and improved joint reception CoMP with uplink PHY level combining. Going from \secrev{Split 0 to Split 3}, more functions are hosted at the CU. In addition to increasing network performance, a higher centralization level can lead to more computing cost savings \cite{vran_murti2}. However, centralizing more functions increases the data load to be transferred to the CU, going from $\lambda$ in \secrev{Split 1 to 2.5 Gbps in Split 3} for each BS, and has a stricter delay requirement. Table \ref{table:splits} summarizes vRAN split options and their requirements\footnote{\secrev{The requirements are tailored from \cite{smallcell,vranmec_andres} by following settings: 1 user per TTI, 20MHz channel bandwidth, 1 carrier component, UE IP MTU 1500 bytes, $2 \times 2$ MIMO.} }.  

%

\begin{table}[t] \centering
	\begin{small}
		\begin{tabular}{@{}lll@{}}\toprule
			\textbf{}& \textbf{Flow (Mbps)} & \textbf{Delay Req. (ms)}  
			\\ \midrule
			{Split 0 \secrev{(S0)} } &      $\lambda$   & $30$          
			\\ \hdashline
			{Split 1 \secrev{(S1)} } &  {$\lambda$} & $30$ 
			\\ \hdashline
			{Split 2 \secrev{(S2)} } &  {$1.02\lambda+1.5$}   & $2$
			\\ \hdashline
			{Split 3 \secrev{(S3)} } &   {$2500$}   & $0.25$ \\ 
			\bottomrule
		\end{tabular}
	\end{small}
	\caption{\small Data and delay requirements of vRAN split when the traffic load is $\lambda$ Mbps \cite{smallcell, vranmec_andres}.}
	\label{table:splits}
	\vspace{-3mm}
\end{table}

\textbf{RAN}. We model a vRAN architecture with a graph $G = (\mathcal{I}, \mathcal{E})$ where $\mathcal{I}$ has a subsets $\mathcal{N}$ of the $N=|\mathcal{N}|$ DUs, \secrev{$\mathcal{L}$ of the $L=|\mathcal{L}|$ routers} and a CU (index $0$). Each node is connected through a link of $(i,j)$ with a set $\mathcal{E}$ of links and has capacity $c_{ij}$ (Mbps) each. The DU-$n$ is connected to $\{0\}$ with a single path (e.g., shortest path) $p_{n0}$; hence, we define $r_{p_{n0}}$ as the amount of data flow (Mbps) to be transferred and routed through a path $p_{n0} \! := \!  \{(n,i_1), ..., (i_k,0) \! : \! (i,j) \! \in \! \mathcal{E} \}$. The BS functions are deployed in servers using virtual machines (VMs)\footnote{\secrev{Each BS
	function can operate as a virtual network function (VNF), and the VNFs can be executed on top of a single VM or multiple VMs}}. Each server has a processing capacity, i.e., $H_n$ for DU-$n$ and $H_0$ for CU. Naturally, a central server has a higher computing performance and capacity, \secrev{hence} $H_0 \! \geq \! H_n$. \secrev{We define} $\rho_{o}^c $ and $ \rho_{o}^d$ as the incurred computational load (cycle/Mb/s) in results of deploying the split configuration $o \! \in \! \{0,1,2,3\}$ at each CU and DU, respectively. 

\textbf{Demand $\&$ Cost}. We focus on the uplink transmission where $\lambda_{n} \geq 0$ (Mbps) is the aggregate data flow of DU-$n$ to serve the users traffic;
hence, there are $N$ different flows in the network. We denote $ \bm{\alpha} = (\alpha_n, n \in \mathcal{N})$ and $\bm{\beta} \!=\! (\beta_n, n \in \mathcal{N})$ as the VM instantiation cost (monetary units) and the computing cost (monetary units/cycle) at the DUs, respectively, while $\alpha_0$ and $\beta_0$ are the respective cost for the CU. We also have a routing cost $\zeta_{p_{n0}}$ (monetary units/Mbps) for each path $p_{n0}$. This cost arises from the network links being leased from third parties or maintaining the links. 

\textbf{Problem Statement.} We have four choices of the \secrev{splits} for each BS in vRANs. What is the best-deployed split for each BS that minimizes the total network cost? The decision leads to interesting problems. Each \secrev{split} generates a different DU-CU data flow and has a distinct delay requirement. Executing more functions at the CU is more efficient in computing cost; however, it produces a higher load for xHaul links. \secrev{The BSs share the same capacitated servers and network links, where each split decision is interdependent. Moreover, the behaviour of the vRAN system (e.g., resources, performance) is complex and highly non-trivial, which makes complete assumptions of the model can be unfeasible or inaccurate. The goal is to design a framework to solve this problem by taking minimal assumptions about the model of the system.}

\vspace{-2mm}
\section{Formalization of vRAN Split Problem} \label{sec:problem}
\vspace{-1mm}
The BS functions can be deployed at the DUs or CU \secrev{depending on} the splits, as seen in Table \ref{table:splits}. \secrev{Each split} must respect to the \emph{chain of functions} $f_0 \!\rightarrow\! f_1 \!\rightarrow\! f_2 \!\rightarrow\! f_3$\footnote{\secrev{$f_0$ is a function that encapsulates RF layers. Then, $f_1, f_2$ and $f_3$ are the functions for Layer 1 (PHY), Layer 2 (MAC, RLC) and Layer 3 (PDCP, RRC and the upper layers), respectively. }}. Thus, we define $x_{on} \in \{0,1\}$ as the decision for deploying split $o \in \{0,1,2,3\}$ at DU-$n$. For instance, $x_{0n} = 1$ is for deploying $f_0, f_1, f_2, f_3$ (Split 0); $x_{1n} = 1$ for $f_0, f_1, f_2$ (Split 1); $x_{2n}= 1$ for $f_0, f_1$ (Split 2); or $x_{3n}= 1$ for $f_0$ (Split 3) at DU-$n$. 
We only deploy a single split configuration for each BS. Therefore, \secrev{a} set of eligible \secrev{splits} is:
%
\begin{align} \label{eq:setx}
\mathcal{X} =  &\Bigl\{  \bm{x}_n  \in \{0,1\} \Big| \sum_{o=0}^3 x_{on}=1 , 
\ \ \forall n \in \mathcal{N}  \Bigr\} , 
\end{align}
where $\bm{x}_n = ( x_{on}, \forall o  )$ and $\bm{x} = (\bm{x}_n, \forall n)$. The BS functions,  $f_1, f_2$ and $f_3$, are deployed \secrev{using} VMs at each server. We have computational processing at the CU and DU-$n$ that must respect \secrev{to} its capacity as:
%
\begin{align} \label{eq:computing1}
\sum_{n \in \mathcal{N}} \lambda_{n} \sum_{o =0}^3    x_{on} \rho^{c}_o \leq \ H_0, 
\end{align}
%
%
\begin{align} \label{eq:computing2} \lambda_{n} \sum_{o =0}^3   x_{on} \rho^{d}_o \leq H_n, \ \forall n \in \mathcal{N}. 
\end{align}
%

\textbf{Data Flow $\&$ Delay.}  Let define $r_{p_{n0}}$ (Mbps) as the amount of data flow (Mbps) to be transferred through a path $p_{n0}$. Hence, the flow must respect \secrev{capacity of each link}:
%
\begin{align} \label{eq:route1}
\sum_{n \in \mathcal{N}} r_{p_{n0}} I^{ij}_{p_{n0}} \leq c_{ij}, \ \ \forall (i,j) \in \mathcal{E},
\end{align}
where $I^{ij}_{p_{n0}} \in \{ 0,1 \}$ indicating whether the link $(i,j)$ is used by path $p_{n0}$. Assuming a single path (e.g., shortest path), the amount of data flow depending on each split configuration is \cite{fluidran_andres}:
%
\begin{align} \label{eq:route2}
r_{p_{n0}} \!=\! \lambda_{n} (x_{0n} + x_{1n}) + x_{2n} (1.02 \lambda_{n} + 1.5) + 2500 x_{3n}.
\end{align}
We let $d_{p_{n0}}$ denote the incurred delay for routing through path $p_{n0}$ from DU-$n$ to the CU. Each split has to satisfy the respective delay requirement (Table \ref{table:splits}):
%
\begin{equation} \label{eq:delay}
x_{on} d_{p_{n0}} \leq d_o^{\text{max}}, \ \ \forall o, \forall n \in \mathcal{N}.
\end{equation}
\subsection{Objective Function}
We aim to minimize the total network cost consisting of the computational \secrev{costs at the DUs and CU} and the routing cost\footnote{In this case, we follow the linear objective cost function similar to the previous studies \cite{fluidran_andres,vranmec_andres}. However, our solution approach does not restrict only to the linear objective. Our approach \secrev{relies on} the \secrev{scalar} reward and penalty as feedback; hence, it can also be tailored to a non-linear objective.}. The needs of computing cost \secrev{for each BS-$n$} at DU-$n$ is:
\begin{align} \label{eq:cost-du}
V_n(\bm{x}_n) = \alpha_n + \beta_n \lambda_{n} \sum_{o=0}^{3}\rho_{o}^{d} x_{on}.
\end{align} 
We also have \secrev{the required} computing cost \secrev{of BS-$n$} at the CU:
\begin{align} \label{eq:cost-cu}
\secrev{V_{n0}(\bm{x}_n)} = \sum_{o=0}^3 x_{on} (\alpha_0 + \lambda_{n} \beta_0 \rho_{o}^{c} ).
\end{align} 
\secrev{The first terms in \eqref{eq:cost-du} and $\eqref{eq:cost-cu}$ represent the required instantiating cost at each DU and CU for BS-$n$. The last terms in \eqref{eq:cost-du} and \eqref{eq:cost-cu} are the required data processing cost by each DU and CU to serve BS-$n$ load.}
Next, we have the cost to route the data flow from DU-$n$ to the CU:
\begin{align} \label{eq:cost-route}
U_{n0}(\bm{x}_n) =  \zeta_{p_{n0}} r_n (\bm{x}).
\end{align} 
Finally, we have the total vRAN cost as:
\begin{align} \label{eq: total-cost}
J(\bm{x}) &= \sum_{n \in \mathcal{N}} \Big(  V_n(\bm{x}_n) + U_{n0}(\bm{x}_n) +\secrev{ V_{n0}(\bm{x}_n}) \Big), \end{align}
which leads to the following problem:
\begin{align}
\mathbb{P}: \,\,\,\, & \underset{\bm{x} \in \mathcal{X}}{\text{minimize}} \   J(\bm{x}), \ \ \ \notag  \text{s.t} \ \  
 (\ref{eq:computing1}) - (\ref{eq:delay}). \notag
\end{align}
$\mathbb{P}$ is a combinatorial problem to decide the function placement $\bm{x}$ for all the BSs and serve the traffic load $\bm{\lambda}$ with DU-CU path $p_{n0}$ for each BS-$n$ in the network graph $G = (\mathcal{I}, \mathcal{E})$. Next, we discuss the complexity of $\mathbb{P}$.

\subsection{Complexity Analysis}

The complexity of $\mathbb{P}$ can be identified from the polynomial reduction of \textit{multiple-choice multidimensional knapsack problem (MMKP).} 

\textbf{MMKP.} Let suppose there are $N$ items with values ${v_1, v_2, ..., v_N}$. We also have $r_1, r_2, ..., r_N$ correspond to the required resources to pick the items. In the 0-1 knapsack problem (KP), the aim is to pick the items $x_i \in \{0, 1\}, \secrev{\forall i}$ that maximize the total value $\sum_{i=1}^N  x_i v_i$, subject to constraint \secrev{$\sum_{i=1}^N x_i r_i \leq R$}. This is a well-known NP hard problem and there is a pseudo-polynomial algorithm using a dynamic programming concept that has complexity $\mathcal{O}(NR)$ \cite{mmkp_convexhull}.  MKKP is a variant of 0-1 KP where there are $M$ groups of items, e.g., group $i$ has $l_i$ items. Each item has a specific value $v_{ij}$ corresponds to $j$-th item of $i$-th group and needs $K$ resources. Hence, each item in a group has a resource vector $\bm{r}_{ij} = (r_{ij1}, ..., r_{ijK} )$ and $\bm{R} = (R_1, ..., R_K)$ is the resource bound of the knapsack. The aim is to exactly pick one item from each group, e.g., $ \sum_{j=1}^{l_i} x_{ij} = 1, x_{ij} \in \{0,1\}$ that maximizes the total value: $\sum_{i=1}^{M} \sum_{j=1}^{l_i} x_{ij} v_{ij}$, subject to the resource constraint: $\sum_{i=1}^{M} \sum_{j=1}^{l_i} x_{ij} r_{ijk} \leq R_k, k = 1,...,K$. 

Finding an exact solution for MMKP is also NP-hard \cite{mmkp_convexhull}. It is also worth noting that the search space for solution in MMKP is smaller than other KP variants; hence, exact solution is not implementable in many practical problems as there is more limitation of picking items from a group in MMKP instance \cite{mmkp_convexhull}. Next, We prove that $\mathbb{P}$ is harder than MMKP. 

\noindent
\begin{theorem} \label{theo:mmkp}
	\textit{MKKP can be reduced to $\mathbb{P}$ in polynomial time, e.g., MMKP $\leq_P \mathbb{P}$ }.
\end{theorem}  

\noindent
\textit{Proof.} Let suppose we have unlimited link capacity, no routing cost and no delay requirement. Hence, all paths of \secrev{the} DU-CU pair are eligible and \eqref{eq:route1}-\eqref{eq:delay} are always satisfied. This problem then can be mapped to MMKP by setting: 1) $M$ groups to $N$ BSs, 2) each $i$-th group with $l_i$ items to each BS-$n$ with $|o|=4 $ of split options, 3) $j$-th item of $i$-th group to the split $o_n$ of BS-$n$, 4) $r_{ij}$ to the \secrev{incurred} computing loads, e.g., $\lambda_{n}\rho_{i}^c$ and $\lambda_{n}\rho_{i}^d$, and 5) the knapsack constraints to computing constraints $H_n$ and $H_0$. The value $v_{ij}$ of item-$j$ in group-$i$ also can be mapped with the costs (e.g., computing and routing) of deploying split-$o$ of BS-$n$, where the MMKP is a maximization problem and $\mathbb{P}$ is a minimization problem.  We can see the reduction is of polynomial time: we select the functional split for every BS correspond to that we activate an item that we pick to a knapsack in each group. Therefore, we can conclude that if we can solve $\mathbb{P}$ in polynomial time we also can solve any MMKP problem. 

%
%

\vspace{-2mm}
\section{Constrained Deep Reinforcement based Functional Split Optimization Framework} \label{sec:solution}

\secrev{We leverage a constrained deep RL paradigm to solve our vRAN splitting problem by treating the vRAN system as a black-box environment, which makes minimal assumptions about the underlying system. Consequently, our RL agent does not need to know the information about the formulation in \eqref{eq:setx}-\eqref{eq: total-cost} to decide the splits. Our agent relies on the scalar reward and penalization returned from the environment to assess the quality of the solutions.}
%
%
%
\secrev{At each episode, our agent observes a \textit{state} of incoming a sequence of all BS functions drawn from the \textit{environment} of vRANs, takes an \textit{action} to decide the splits for all the BSs, and expects to receive feedback signals of the \textit{reward} (total network cost) and \textit{penalization} (for violating the constraints). Our state comprises of a sequence information of BS functions: $\mathcal{F} = \{ \mathcal{F}_n \}_{n=1}^{N}$,  where $\mathcal{F}_n$ is a set of BS-$n$ functions. Given the input state, our agent assigns $\mathcal{O} \!\!=\!\! \{o_n \! \in \! \{0,1,2,3\}, \forall n \in \mathcal{N} \}$ as a set of selected splits for all the BSs, which decides the placement of BS functions at the CU or DUs. Our objective is to minimize the total network cost while enforcing the constraint requirements. Given the selected splits, our agent expects to receive scalar values from the environment consisting of: \textit{i)} $J(\mathcal{O}|\mathcal{F})$, the total induced cost and \textit{ii)} $\xi(\mathcal{O}|\mathcal{F})$, the weighted sum of penalization. Further, we consider a particular RL algorithm using one-step constrained policy optimization and neural network architecture, where the interactions are narrowed to a single time step at every episode, and our agent learns iteratively over episodes.}


\secrev{Our goal is to design a stochastic policy $\pi_\theta(\mathcal{O}|\mathcal{F})$ parameterized by a neural network with weights $\theta$ to predict the splits for all the BSs to minimize the total cost while satisfying constraint requirements. However, we have the $N$ BSs that need to deploy the splits together, where each has four possible split options. Each split decision is also interdependent as the BSs share the same network links and computing servers. Consequently, our problem has a combinatorially large discrete action space with a total of $4^N$ possible actions. Such a curse dimensionality in high dimensional spaces can be avoided by modelling complicated joint probability distributions using the chain rule decomposition. Therefore, we design our policy based on a chain rule by factorizing the output probability, parameterized by a neural network with weights $\theta$ as:
	\begin{align}
		\pi_\theta(\mathcal{O}| \mathcal{F}) = \prod_{n=1}^{N} \pi_\theta(o_n| o_{(<n)}, \mathcal{F}_n). 
	\end{align}
	This policy strategy assigns a higher probability to the splits for having a lower cost and vice versa for every BS, which also can be represented by individual softmax modules (e.g., at the output layer). Motivated by \cite{seq2seq, attention_bahdanau} that uses neural networks \thirdrev{to estimate} the same factorization \thirdrev{of our stochastic policy} for machine translation, we design our policy network using an encoder-decoder sequence-to-sequence model based on LSTM networks. Our policy network architecture, which also utilizes an attention mechanism, captures the dependency and correlation between split decisions. This architecture allows our policy to read input information from all BS functions, then maps them into split selections for all the BSs.
	%
	In the training, we use a batch of $B$ i.i.d samples on the stochastic policy to select the splits and generate several pretraining models. In the test, we perform an inference through a search strategy by greedy decoding or temperature sampling.
	}

\vspace{-2mm}
\subsection{\secrev{Policy} Network Architecture}
Our policy network infers a strategy to deploy the splits for all the BSs, given a sequence information of BS functions as an input $\mathcal{F} = \{\mathcal{F}_1, ...., \mathcal{F}_N \}$. It is constructed from an encoder decoder sequence-to-sequence model with an attention mechanism based on LSTM networks \cite{seq2seq,attention_bahdanau}.
\secrev{We also consider a batch training by drawing} a batch of $B$ i.i.d samples with different sequence order \secrev{to encourage the exploration further}. 

\begin{figure}[t!] 
		\centering
		\includegraphics[width=0.45 \textwidth]{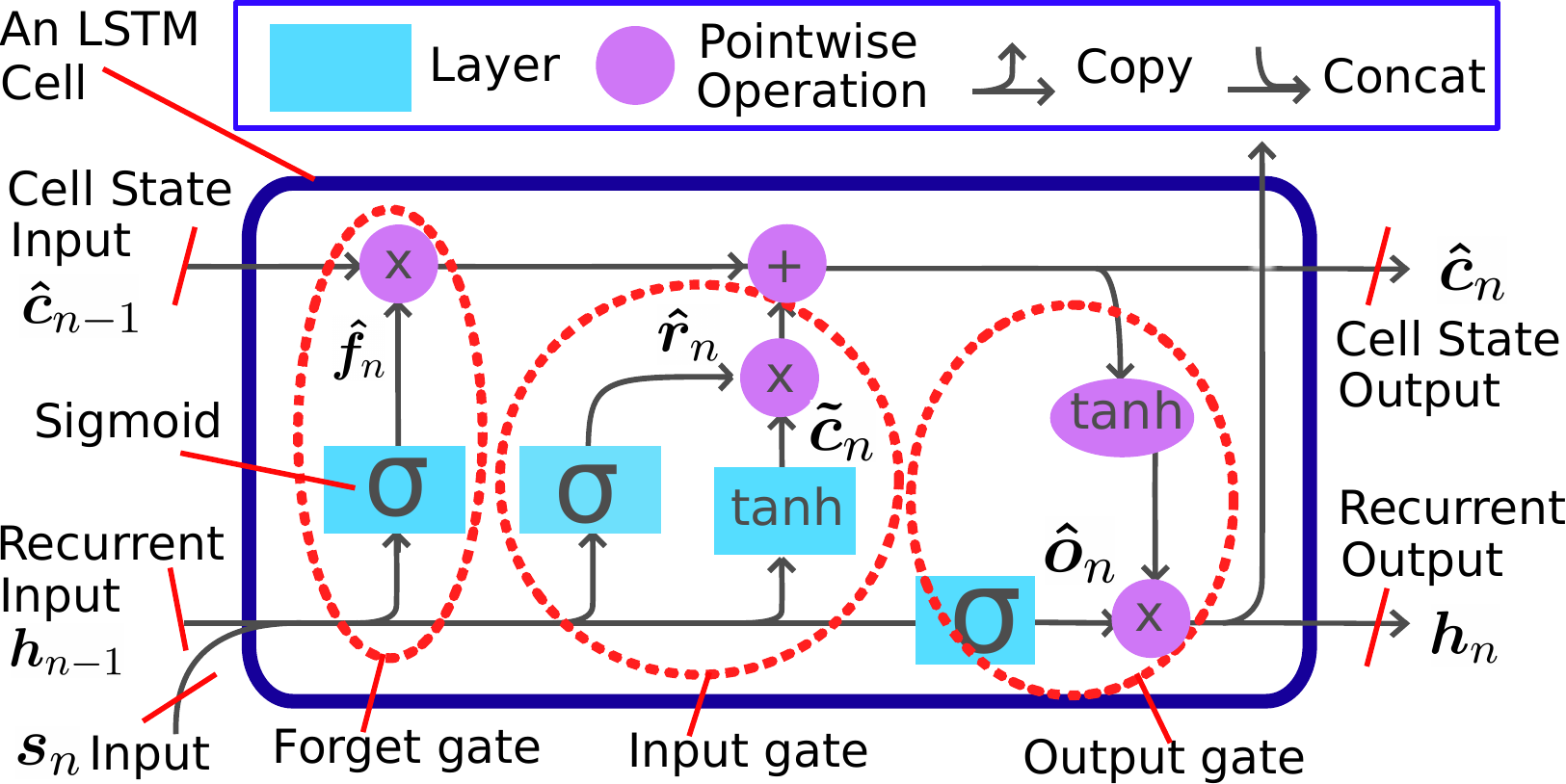}   
		\caption{\small \secrev{A generic architecture of an LSTM cell.}  } 
		\label{fig:lstm}
		\vspace{-3mm}
	\end{figure}

\secrev{\textbf{LSTM structure.} We leverage LSTM networks, a particular RNN architecture \cite{lstm},  to construct  our sequence-to-sequence model that maps the input BS functions into split decisions for all the BSs.
	An LSTM cell has three main structures comprising of: \textit{(i)} a forget gate that receives the cell state input and learns how long should memorize or forget from the past; \textit{(ii)} an input gate that aggregates the current input and the output of past steps, then feeds them to the activation function; and \textit{(iii)} an output gate that provides the LSTM output from the combination of current cell state and the output of input gate. The relationship of these blocks can be expressed as:
	\begin{align}
		&\bm{\hat{f}}_n = \sigma \big( W_f \big[\bm{h}_{n-1}^T, \bm{s}_{n}^T  \big]^T + \bm{b}_f \big), \\
		&\bm{\hat{r}}_n = \sigma \big( W_r \big[\bm{h}_{n-1}^T, \bm{s}_{n}^T   \big]^T + \bm{b}_r \big), \\
		&\bm{\tilde{c}}_n = \tanh \big( W_c \big[\bm{h}_{n-1}^T, \bm{s}_{n}^T   \big]^T + \bm{b}_c  \big), \\
		&\bm{\hat{c}}_n = \bm{\hat{f}}_n *  \bm{\hat{c}}_{n-1} + \bm{\hat{r}}_n  * \bm{\tilde{c}}_n, \\
		&\bm{\hat{o}}_n = \sigma \big( W_o \big[\bm{h}_{n-1}^T, \bm{s}_{n}^T   \big]^T + \bm{b}_o  \big), \\
		& \bm{h}_n = \bm{\hat{o}}_n * \tanh(\bm{\hat{c}}_n),
	\end{align}
	where function $\sigma(x) \triangleq \frac{1}{1 + \exp(-x)} $ is the sigmoid function and symbol $*$ is element-wise multiplication. The weight and bias matrices for the respective forget, input and output gates of the LSTM cell are represented by $W_f, W_r, W_c, W_o$  and $\bm{b}_f, \bm{b}_r, \bm{b}_c, \bm{b}_o$.
	Multiple LSTM layers can be further stacked one on top of another (a stacked LSTM) to create a deeper model, which may obtain more accurate prediction. Each LSTM cell reads an input of embedding vector representation $\bm{s}_n \in [ -1,1]^E$ translated from each input $\mathcal{F}_n$, where $E$ is the embedding size.
	The structure of an LSTM cell is illustrated in Fig \ref{fig:lstm} and utilized to construct our sequence-to-sequence model.}
	%
%

\begin{figure}[t!] 
	\centering
	\includegraphics[width=0.49 \textwidth]{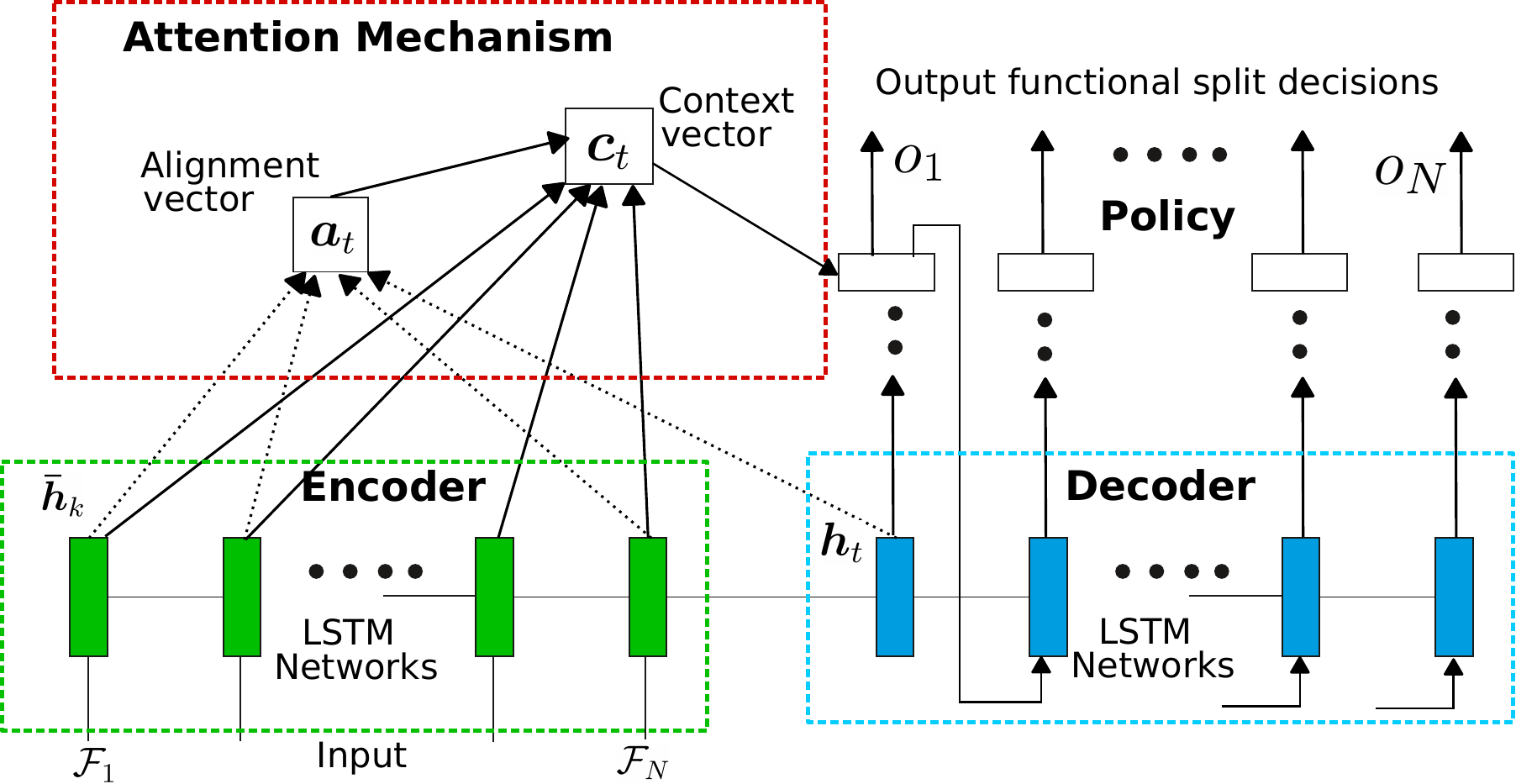}   
	\caption{\small \secrev{\textbf{Policy Network.} CDRS utilizes a neural network architecture to approximate the stochastic policy over the solution. It is constructed by an encoder-decoder sequence-to-sequence model with attention mechanism based on LSTM networks.} }
	\label{fig:neural}
	\vspace{-3mm}
\end{figure}

\secrev{\textbf{Policy Network.}} Our policy network is built from an encoder-decoder sequence-to-sequence model based on LSTM networks. \secrev{One main drawback of vanilla sequence model is generally unable to learn accurately long sequence. Therefore, the vanilla model may not be able learn our problem with large number of BSs.
An attention mechanism comes to address this issue as it considers all the hidden state from all input sequences.} The encoder read the entire input sequence to a fixed-length vector. The decoder decides \secrev{the deployed split of each BS} at each step from an output function based on its own previous state combined with an attention over the encoder hidden states \cite{attention_bahdanau}. The decoder network hidden state is defined with a function: $\bm{h}_t = f(\bm{h}_{t-1}, \bm{\bar{h}}_{t-1}, \bm{c}_t)$, \secrev{where $\bm{c}_t$ and $\bar{\bm{h}}_{t}$ are the context vector and the source hidden state at time step $t$}. 
Our model derives the context vector $\bm{c}_t$ that captures relevant source information that helps to predict the splits. The main idea is to use \secrev{an attention mechanism}, where the context vector $\bm{c}_t$ takes consideration of all the hidden states of the encoder and the alignment vector $\bm{a}_{t}$: 
\begin{align}
	\bm{c}_t = \sum_{k \in \mathcal{N}} \bm{a}_{tk} \bm{\bar{h}}_k.
\end{align}
Note that the alignment vector has an equal size to the number of steps in the source side, which can be calculated by comparing the current target hidden state of decoder $\bm{h}_t$ with each source hidden state $\bm{\bar{h}}_k$ as:
\begin{align} \label{eq:softmax}
	 \bm{a}_{tk} = \frac{\exp(\text{score}(\bm{h}_t,\bm{\bar{h}}_k))}{\sum_{k'=1}^{N} \exp(\text{score}(\bm{h}_t,\bm{\bar{h}}_k')))}
\end{align}
This \secrev{alignment} model gives a score $\bm{a}_{tk}$ which describes how well the pair of input at position $k$ and the output at position $t$. The \secrev{alignment} score is parameterized by a feed-forward network where the network is trained jointly with the other models \cite{attention_bahdanau}. The score function is defined by a non-linear activation function following Bahdanau's additive style:
\begin{align}\label{eq:score}
	\text{score}(\bm{h}_t,\bm{\bar{h}}_k) = \bm{v}_a^{\top} (\tanh(\bm{w}_1 \bm{h}_t +  \bm{w}_2 \bm{\bar{h}}_k )),
\end{align}
where $\bm{v}_a^{\top} \in \mathbb{R}^{n}, \bm{w}_1 \!\in\! \mathbb{R}^{n \times n} $ and $\bm{w}_2 \!\in\! \mathbb{R}^{n \times n}$ are \secrev{defined as the weight matrices to be learned in the alignment model, and $n$ is the size of hidden layers}. The overall architecture of our policy network is illustrated in Fig. \ref{fig:neural}.
\vspace{-2mm}
\subsection{Constrained Policy Gradient with Baseline}
%
%
%
%
%
%
\secrev{We train the above neural network model using a constrained policy gradient method with a self competing baseline.}
We define the objective of $\mathbb{P}$ as an expected reward that is obtained for every vector of weights $\theta$. Hence, the expected cost $J$ in associated with the selected split $o_n$ given BS-$n$ functions \secrev{is denoted as}:
\begin{align}
	J^\pi(\theta|\mathcal{F}_n) = \underset{o_n \sim \pi(.|\mathcal{F}_n) }{\mathbb{E}} [ J(o_n) ],
\end{align}
and we have the expected of total cost from all BSs:
\begin{align} \label{eq:total_cost_theta}
J^\pi(\theta) = \underset{o_n \sim \mathcal{O} }{\mathbb{E}} [ J(\theta|\mathcal{O}) ].
\end{align}
The vRAN system has constraints of delay requirement and computational and link capacity. 
%
%
Therefore, our original problem turns to a primal problem as:
\begin{align} 
\mathbb{P}_{1\text{P}}: \ \underset{\pi \sim \Pi }{\text{min}} \  J^\pi(\theta); \ \ \text{s.t.} \ \secrev{J_{C_i}^\pi(\theta) \leq 0, \forall i}, \notag
\end{align}
where \secrev{we define $J_{C}^\pi(\theta) = \big(J_{C_{i}}^\pi(\theta), \forall i \big)$} as a function of constraint dissatisfaction to capture the penalization that the environment returns for violating each $i$ constraint requirement, e.g., computing, link, delay. 
In this problem, we consider parametrized stochastic policy using a neural network. In order to ensure the convergence of our policy to constraint \secrev{satisfaction}, we follow \cite{reward_constraint} and make assumptions:
\begin{assumption} \label{assumption:cost}
	$J^\pi$ is bounded for all policies $\pi \in \Pi$.
\end{assumption}
\begin{assumption} \label{assumption:localminima}
	Each local minima of $J_{C}^\pi(\theta)$  is a feasible solution.
\end{assumption}
\noindent
Assumption \ref{assumption:localminima} describes that any local minima $\pi_\theta$ satisfies all constraints, e.g., $J_{C_i}^\pi(\theta) \leq 0, \forall i$. It is the minimal requirement that guarantees the convergence of a gradient algorithm to a feasible solution. The stricter assumptions, e.g., convexity, may guarantee the optimal solution.
%

Next, we reformulate $\mathbb{P}_{1\text{P}}$ to unconstraied problem with Lagrange relaxation method \cite{Bertsekas}. The penalty signal is also included aside \secrev{from the} original objective for infeasibility, which leads to a sub-optimality for infeasible solutions. Given $\mathbb{P}_{\text{1P}}$, we have the dual function:
\begin{align} 
	\label{eq:dual_function}
	g({\mu}) = \underset{\theta }{\text{min}} \  J_L^\pi({\mu},\theta) &= \underset{\theta }{\text{min}} \  J^\pi(\theta) + \sum_{i} \mu_i J_{C_{i}}^\pi(\theta) \notag \\
	&=\underset{\theta }{\text{min}} \  J^\pi(\theta) + J_\zeta^\pi(\xi),
\end{align}
where $\mu \!=\! (\mu_i, \forall i), J_L^\pi({\mu},\theta)$ and $J_\zeta^\pi(\xi)$ are the  penalty coefficients (Lagrange multipliers), Lagrange objective function and the expected penalization, respectively. Then, we define the dual problem:
\begin{align}
\mathbb{P}_{1\text{D}}: \ \underset{{\mu} }{\text{max}} \  g({\mu}). \notag
\end{align}
$\mathbb{P}_{1\text{D}}$ aims to find a local optima or a saddle point $(\theta({\mu}^*), {\mu}^*)$, which is a feasible solution. The feasible solution is a solution that satisfies:  $J_{C_{i}}^\pi(\theta) \leq 0, \forall i$. 
To compute the weights $\theta$ that optimize the objective, we use Monte-Carlo policy gradient and stochastic gradient descent by the following update:
\begin{align}\label{eq:update1}
	\theta_{k+1} = \theta_{k} - \eta_a(k) 	 \nabla_\theta J_L^\pi({\mu},\theta), 
\end{align}
where $\eta_a(k) $ is the step-size. The gradient $\nabla_\theta J_L^\pi({\mu},\theta)$  with regards to weights $\theta$ can be calculated using a log-likelihood method as:
\begin{align}
\nabla_\theta J_L^\pi(\theta) = \underset{\mathcal{O} \sim \pi_\theta(.|\mathcal{F}) }{\mathbb{E}} [ L(\mathcal{O}|\mathcal{F}) \ \nabla_\theta \log \pi_\theta(\mathcal{O}|\mathcal{F}) ].
\end{align} 
$L(\mathcal{O}|\mathcal{F})$ represents the total cost with penalization obtained from: 
	 $L(\mathcal{O}|\mathcal{F}) = J(\mathcal{O}|\mathcal{F}) + \xi (\mathcal{O}|\mathcal{F}) $, where
$J(\mathcal{O}|\mathcal{F})$ is the total network cost in each iteration and $\xi (\mathcal{O}|\mathcal{F}) = {\mu} C(\mathcal{O}|\mathcal{F})$ is the weighted sum of constraint dissatisfaction of $C(\mathcal{O}|\mathcal{F})$. 

The penalty coefficient ${\mu}$ is set manually \cite{vnf_drl_solozabal,cmpd_solo} for CDRS-Fixed within a range $[0, \mu_{\text{max}} ]$\footnote{If Assumption \ref{assumption:localminima} is satisfied, $\mu_{\text{max}}$ can be set to $\infty$ \cite{reward_constraint}.}. In this case, the selection of ${\mu}$ can be set following intuition approach in \cite{vnf_drl_solozabal} (Appendix C), i.e., agent will not pay attention to penalty if $\mu = 0$, and it will only converge to penalization if $\mu = \infty$. Hence, selecting the appropriate penalty coefficient is important in this case. Otherwise, we can follow a less intuitive approach by adaptively updating the penalty coefficient (CDRS-Ada). CDRS-Ada is updated based on the primal-dual optimization (PDO) method inspired from \cite{pdo_risk}. Hence, we update the penalty coefficient in the ascent direction as:
\begin{align} \label{eq:update2}
{\mu}_{k+1} &= {\mu}_{k} + \eta_d(k) \nabla_\mu J_L^\pi({\mu},\theta) \\
& = {\mu}_{k} + \eta_d(k) (  J_{C}^\pi(\theta))_+, 
\end{align} 
where $\eta_d(k)$ is the step-size (Dual) and \secrev{$\nabla_\mu J_L^\pi({\mu},\theta) = \mathbb{E}_{\mathcal{O} \sim \pi_\theta(.|\mathcal{F})} [ C(\mathcal{O}|\mathcal{F}) ]$ is the gradient with respect to $\mu$}. The penalty coefficient ${\mu}_{k}$ is updated for every $k$-th iteration and will converge to a fixed value once the constraints are satisfied \cite{reward_constraint,pdo_risk}. 
Then, Monte-Carlo sampling can be applied to approximate \secrev{$J_L^\pi(\theta)$} by drawing  $B$ i.i.d samples $ \!\mathcal{F}^1,...,\mathcal{F}^B \!\sim\! \mathcal{F}$, which can be written:
\begin{align} \label{eq:lag_grads}
\!	\nabla_{\!\theta} J_L^\pi(\theta) \! \approx \! \frac{1}{B} \! \sum_{i=1}^{B} \! \! \Big(\! L(\mathcal{O}^i | \mathcal{F}^i) \! - \! b_{\theta_v}(\mathcal{F}^i)\! \Big) \! \nabla_{\!\theta} \! \log \! \pi_\theta(\mathcal{O}^i | \mathcal{F}^i), \!\!
\end{align} 
\secrev{where $b_{\theta_v}(\mathcal{F}^i)$ is the baseline estimation given the state input of $i$-th batch, parameterized by a neural network structure with weights $\theta_v$.}

\begin{figure}[t!] 
	\centering
	\includegraphics[width=0.24 \textwidth]{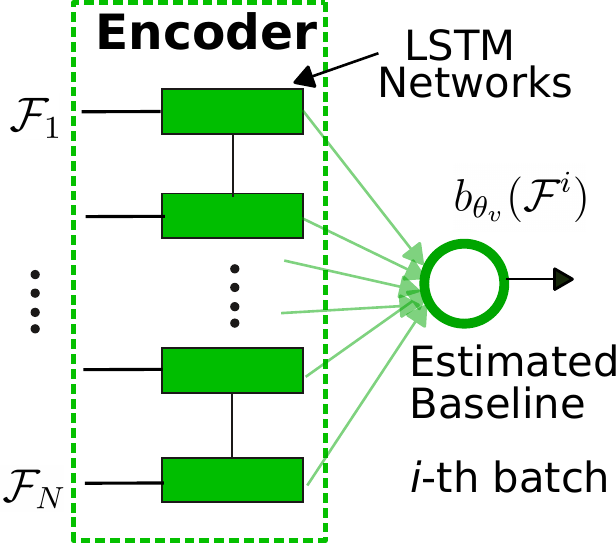}   
	\caption{\small\secrev{\textbf{Baseline Estimator.} The self-competing baseline of CDRS is estimated using an auxiliary network constructed from an LSTM encoder connected to a multilayer perceptron output linear layer.}  } 
	\label{fig:baseline}
\end{figure}

\textbf{Baseline estimator.} The baseline choice can be from an exponential moving average of the reward over time that captures the improving policy in training. Although it succeeds in the Christofides algorithm, it does not perform well because it can not differentiate between inputs \cite{neural_bello}. To this end, we use a parametric baseline $b_{\theta_v}$ to estimate the expected total cost with penalization that typically improves the learning performance. \secrev{We estimate the baseline through} an auxiliary network built from an LSTM encoder connected to a multilayer perceptron output layer. \secrev{The auxiliary network (parameterized by $\theta_v$) that approximates the expected cost with penalization from input $\mathcal{F}$ is trained with stochastic gradient descent.} It employs a mean squared error (MSE) objective, calculated from the prediction of $b_{\theta_v}$ and the total cost with penalization $L(\mathcal{O}^i | \mathcal{F}^i)$, and sampled by the most recent policy (obtained from the environment). We formulate \secrev{the auxiliary network goal is to minimize the below loss function:}
\begin{align} \label{eq:aux_mse}
	\mathcal{L}(\theta_v) = \frac{1}{B} \sum_{i=1}^{B} \left\| b_{\theta_v}(\mathcal{F}^i) - L(\mathcal{O}^i | \mathcal{F}^i) \right\|_2^2.
\end{align}
Fig. \ref{fig:baseline} illustrates the architecture of the auxiliary network for estimating the baseline. 
\begin{figure}[t!] 
	\centering
	\includegraphics[width=0.49 \textwidth]{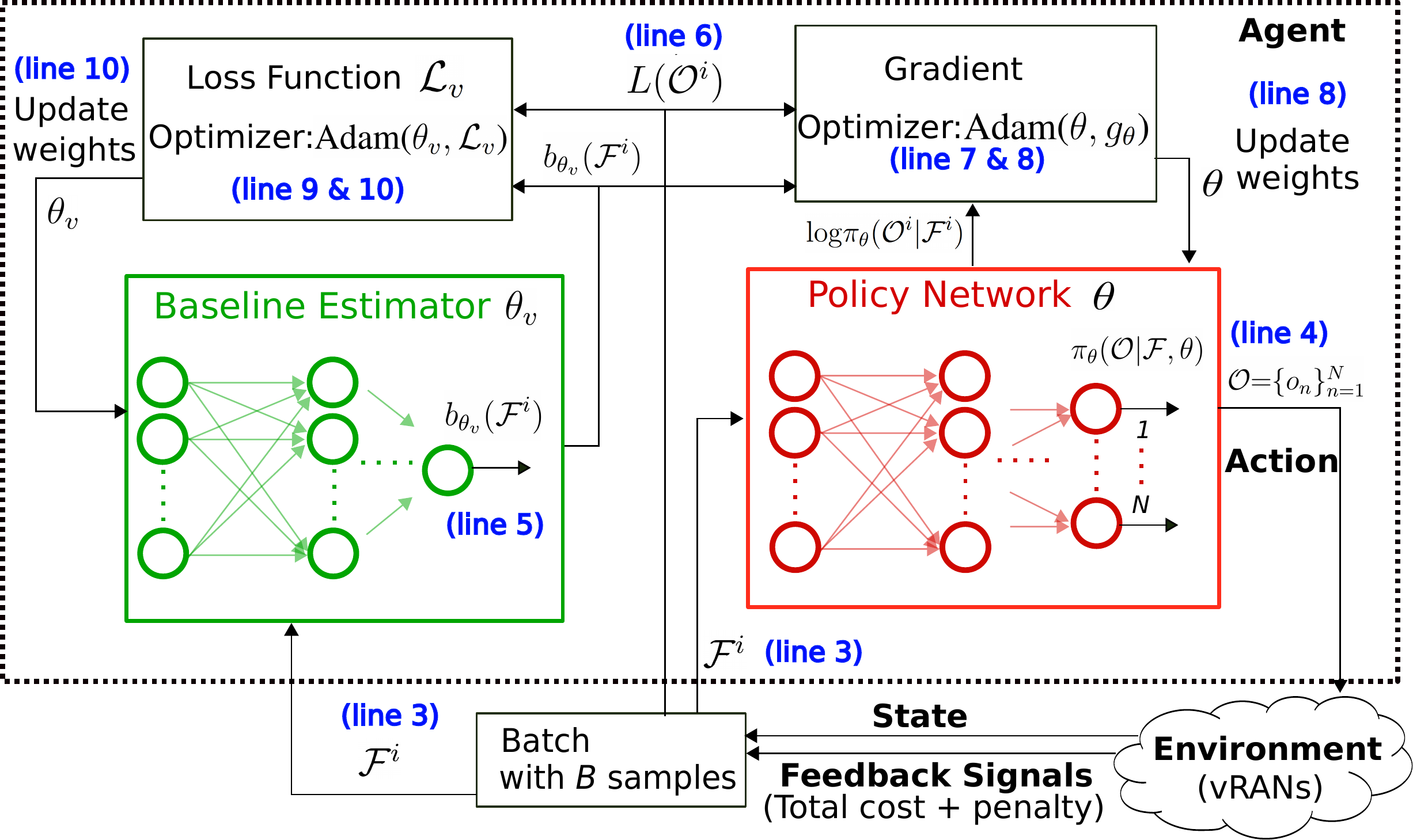}   
	\caption{\small\textbf{CDRS Diagram.} CDRS is trained using a single time step Monte-Carlo policy gradient algorithm, where at every epoch, the interactions with the environment are narrowed to a single time step. Our agent learns the policy iteratively over epochs.} 
	\label{fig:rl_diagram}
	\vspace{-3mm}
\end{figure}
\begin{algorithm}[t!]  \caption{CDRS Training}
	\label{algo:cdrs}
	\SetAlgoLined
	\DontPrintSemicolon
	\KwInput{$K$ (Num of epoch), $B$ (Batch size), $\mathcal{F}$ (Learning set)}
	\KwInitialize{ assign agent and critic (baseline) networks with random weights $\theta$ and $\theta_v. \;$} 
	\For{ $ k=1, ..., K$}  
	{
		$d\theta$ $\leftarrow$ 0 \% Reset gradient \\
		$\mathcal{F}^i \sim $ \text{SampleInput} $(\mathcal{F})$ for $i \in \{1,...,B \}$. \;
		$\mathcal{O}^i \sim $ SampleSolution $(\pi_\theta(.|\mathcal{F}))$ for $i \in \{1,...,B \}$. \;
		$b^i \leftarrow b_{\theta_v} (\mathcal{F}^i)$ for $i \in \{1,...,B \}$. \;
		Compute $L(\mathcal{O}^i)$ for $i \in \{1,...,B \}$. \;
		$g_\theta \leftarrow \frac{1}{B} \! \sum_{i=1}^{B} \! \! \Big(\! L(\mathcal{O}^i) \! - \! b^{i}\! \Big) \! \nabla_{\!\theta} \! \log \! \pi_\theta(\mathcal{O}^i | \mathcal{F}^i)$ from \eqref{eq:lag_grads}. \;
		$\theta \leftarrow$ Adam($\theta, g_\theta$) \%Run Adam algorithm \;
		$\mathcal{L}_v \leftarrow \frac{1}{B} \sum_{i=1}^{B} \left\| b^{i} - L(\mathcal{O}^i) \right\|_2^2 $ from \eqref{eq:aux_mse}. \;
		$\theta_v \leftarrow$ Adam($\theta_v, \mathcal{L}_v$) \%Run Adam algorithm \;
		{\color{black} Update ${\mu}$ from \eqref{eq:update2} \%CDRS-Ada\\ }
		{\color{black} Set ${\mu} = \max(0,{\mu})$ \%CDRS-Ada}
	}
	\Return $\theta, \theta_v, \mu$
	\;
\end{algorithm}

\secrev{To sum up, our training procedures are summarized in Algorithm \ref{algo:cdrs} and illustrated in Fig. \ref{fig:rl_diagram}, which run iteratively by $K$ episodes (epochs) based on a single time-step Monte-Carlo policy gradient with a baseline estimator.}
The sequence of policy updates will converge to a locally optimal policy and the penalty coefficient updates (e.g., CDRS-Ada) will converge to a fixed value when all constraints are satisfied; see also \cite{pdo_risk,reward_constraint}.

\vspace{-1mm}
\subsection{Searching Strategy}
At the test time, evaluating the total network cost is inexpensive \secrev{as it only requires a forward pass from the policy network to decide the splits}. Our agent can add a search procedure during the inference process by considering solution candidates from multiple \secrev{pretraining} models to select the splits. It can help to reduce the inferred policy suffering from a severe suboptimality. 
We employ two different search strategies by greedy decoding and \secrev{temperature sampling} \cite{neural_bello}.

\textbf{Greedy decoding.} It is the simplest search strategy. The idea is to \secrev{greedily} select the splits with the \secrev{highest} probability for having the lowest cost \secrev{from multiple pretraining models during the inference time}. 
Then, we can extend CDRS to CDRS-Fixed-G, which uses a fixed penalty coefficient with greedy decoding and CDRS-Ada-G that uses an adaptive penalty coefficient with greedy decoding. 

\textbf{Temperature sampling.} This method samples through stochastic policy for \secrev{each pretraining model to generate several candidate solutions} then \secrev{decides} the splits with the lowest total cost \secrev{among them} \cite{neural_bello,vnf_drl_solozabal}. As opposed to the heuristic solvers, it does not sample the different split options. Instead, \secrev{it samples through the stochastic policy} and controls the sparsity of the output distribution \secrev{by} a temperature hyperparameter $T$. The softmax function in \eqref{eq:softmax} is modified to $\bm{a}_{tk} = \frac{\exp\big( \text{score}(\bm{h}_t,\bm{\bar{h}}_k)/T \big)}{\sum_{k'=1}^{N} \exp\big(\text{score}(\bm{h}_t,\bm{\bar{h}}_k')/T ) \big)}$ (softmax temperature). \secrev{In the training}, the temperature hyperparameter $T$ is \secrev{set} to 1. \secrev{Meanwhile, we modify to $T>1$ during the test}, hence the output distribution becomes less step, \secrev{which} prevents the model from being overconfident. With this method, we can extend CDRS to CDRS-Fixed-T (fixed penalty coefficient, temperature sampling) and CDRS-Ada-T (adaptive penalty coefficient, temperature sampling). \secrev{Note that this method requires additional time, which depends on the number of samples.}


\vspace{-2mm}

\section{Results and Discussion} \label{sec:results}
\vspace{-1mm}
In this section, we conduct several experiments to evaluate our approach using synthetic and real network datasets. We aim to examine our approach in regards to: \textit{(i)} the behaviour during the training process, \textit{(ii)} the accuracy and solution distributions to the optimality with different penalty coefficient and search strategy settings, \textit{(iii)} the impact of routing costs and traffic loads on the optimality performance and total network cost, and \textit{(iv)} the computational time.

\vspace{-2mm}
\subsection{Environment \& Experiment Setup}
\begin{figure}[t]
	\centering
	\begin{subfigure}[t]{.23\textwidth}
		\centering
		\includegraphics[width=\textwidth]{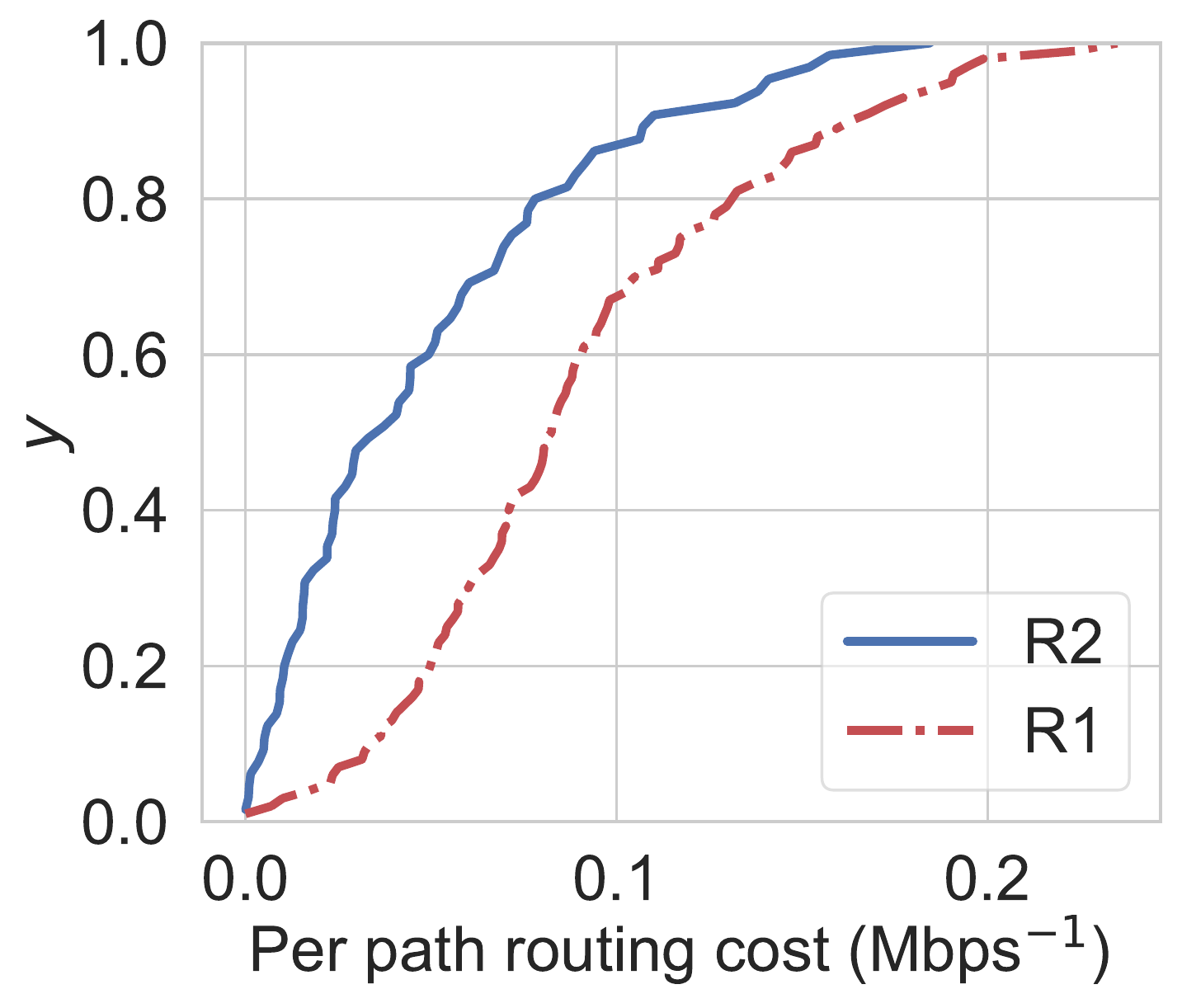}
		\small\caption{\small}
	\end{subfigure}
	\begin{subfigure}[t]{.235\textwidth}
		\centering
		\includegraphics[width=\textwidth]{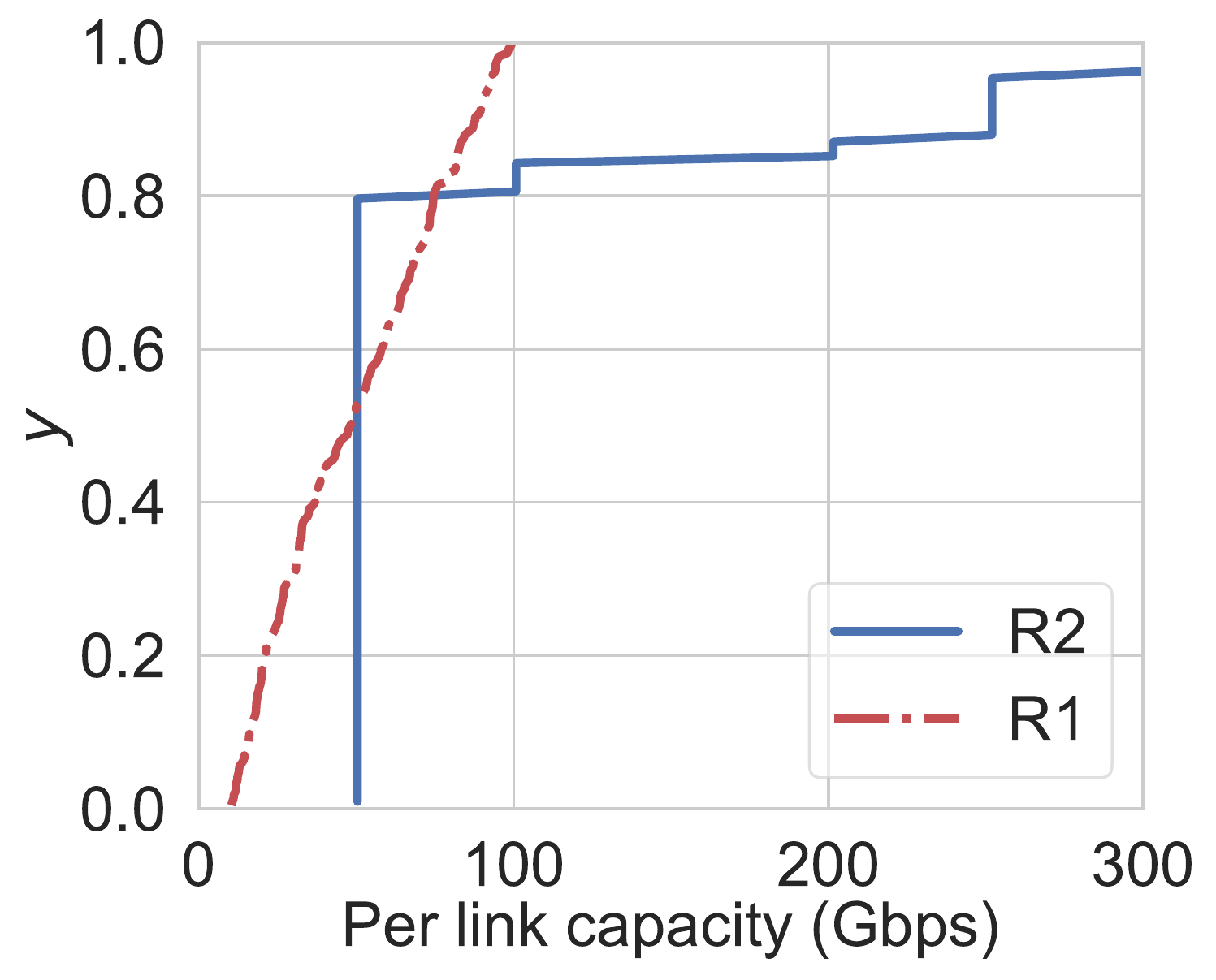}
		\small\caption{\small }
	\end{subfigure}
	\begin{subfigure}[t]{.235\textwidth}
		\centering
		\includegraphics[width=\textwidth]{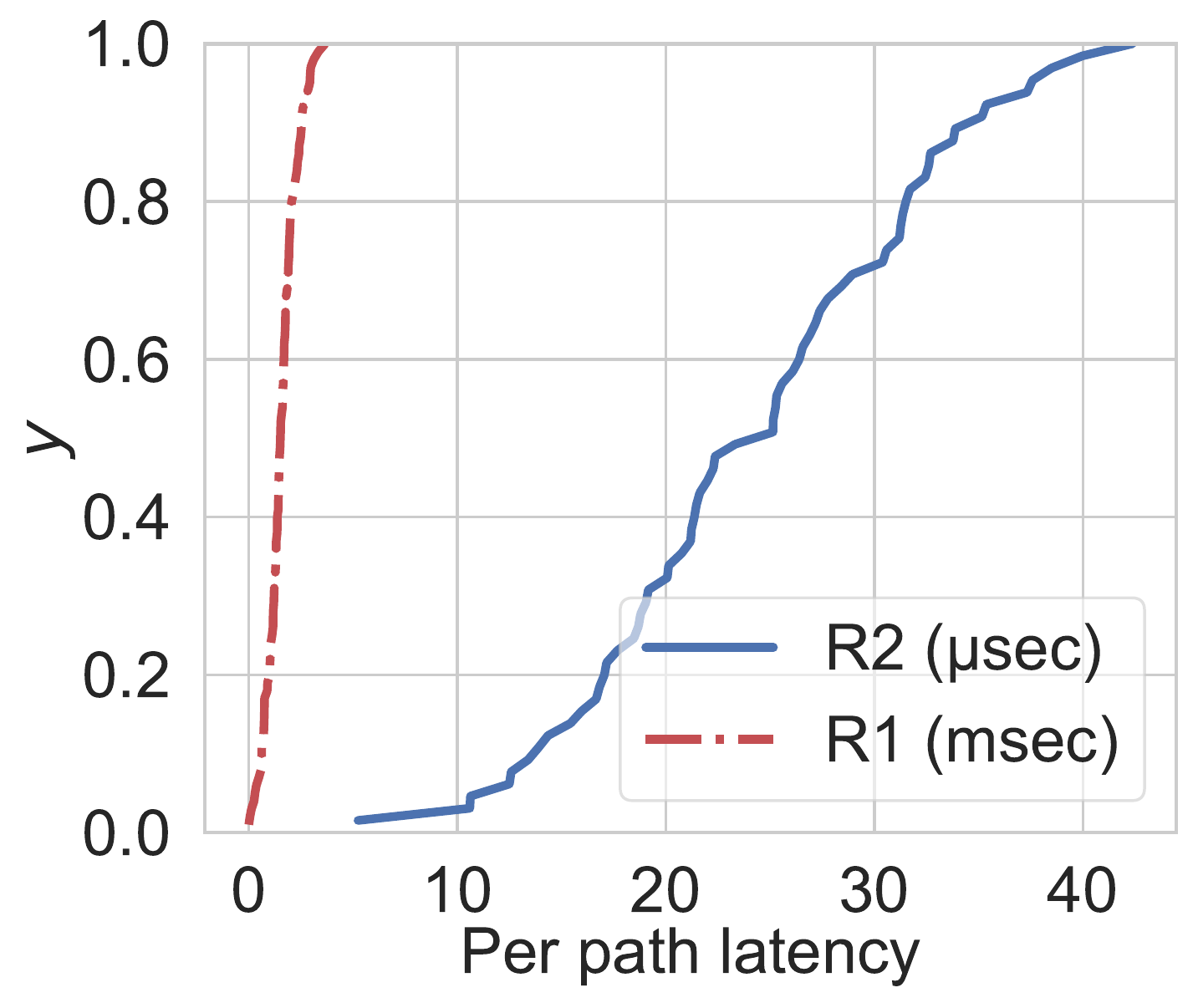}
		\small\caption{\small }
	\end{subfigure}
	\caption{\small \textbf{RANs dist.} eCDF of (a) per-path routing cost, (b) per-link capacity, (c) per-path latency for R1 and R2.}
	\label{fig:ran}	
	\label{fig:ran_params}
	\vspace{-3mm}	
\end{figure}

We use synthetic (R1) and real (R2) network datasets to evaluate our approach. We generate R1 \thirdrev{with} stricter constraints and a larger scale environment than R2. R1 is generated using the Waxman algorithm \cite{waxman} with parameters such as link probability ($\alpha$) and edge length control ($\beta$). These respective parameters $(\alpha,\beta)$ are set to $(0.5, 0.1)$. R1 has 1 CU and 99 DUs. In the case of R2, we utilize a real network dataset from \cite{network_sndb}, which has 1 CU and 63 DUs. \secrev{We assume that the routers are co-located with the DUs.} R1 and R2 differ in parameters, e.g., location, link capacity, weighted link, delay. We use  a standard store-and-forward model to calculate the delay. It is from $12000/c_{ij}$, $4 \mu\text{secs}$/Km and $5 \mu\text{secs}$ for transmission, propagation and processing delay, respectively; see \cite{vranmec_andres}. The link capacity varies to $100$ Gbps (R1) and $252$ Gbps (R2). The path delay reaches to $3658 \ \mu s$ (R1) and $42 \ \mu s$ (R2). In R1, the routing cost per path is calculated from the total cost per link (randomly generated) which belongs to the selected path. A link with a routing cost of 1 monetary unit per Mbps means having the same cost as a DU computing cost. We consider the routing cost within a range of $0.001 - 0.01$ times of DU computing cost (for the same network load) for each link in R1. In R2, we calculate the distance between nodes based on its geolocation dataset from \cite{network_sndb} and charge the cost of $0.01$ monetary units per Mbps/km. Fig. \ref{fig:ran_params} depicts the parameter distributions of our RANs with eCDF.


In this experiment, all system parameters correspond to testbed measurements of previous studies \cite{crancomplexity, vranmec_andres, vran_murti2,cost_vm}. We assume a high load scenario $\lambda_{n} = 150$ Mbps for every DU. This setting is based on 1 user/TTI, $2 \times 2$ MIMO, 20 Mhz (100 PRB), 2 TBs of 75376 bits/subframe and IP MTU 1500B. We use an Intel Haswell i7-4770 3.40GHz CPU as the \textit{reference core}, and set the maximum computing capacity to 75 RCs for CU and 7.5 RCs for each DU. Each split $o \in \{ 0,1,2,3 \}$ inccurs computational load $\rho_{o}^{{d}} = \{ 0.05, 0.04, 0.00325, 0\}$ RCs per Mbps at each DU and $\rho_{o}^{{c}} = \{0, 0.001, 0.00175, 0.05 \} $ RCs per Mbps at the CU. The VM instantiation cost at the CU is half of the DU $(\alpha_0 = \alpha_n/2)$ and the processing cost is set to $\beta_0 = 0.017 \beta_n$. 

Our learning rate is initially set to $\eta_a = 0.0001$ (Agent) and $\eta_b = 0.005$ (Baseline) with the batch size: 128. Our neural network has the number of layers, hidden dimension and embedding size with $1, 32$ and $ 32$, respectively. The temperature hyperparameter is set to $T=1$ by default, so the model computes the softmax function directly. We scale all the original values of weighted paths and traffic loads randomly with uniform distribution $[0,1]$ as in \cite{neural_bello}. Then, we generate three models (RL-pretaining) as outputs of our training with 50000 (in R1) and 15000 (in R2) epochs each. CDRS-Fixed uses a fixed penalty coefficient with $\mu_i =1, \forall i$ for all epochs while CDRS-Ada is set with initial penalty coefficient $\mu_i (0) =1, \forall i$ and step-size $\eta_d = 0.001$. The training is performed with Tensorflow 1.15.3 and Python 3.7.4. In the test, the temperature sampling method uses $16$ samples and $T = 15 $ (softmax temperature). 

\vspace{-2mm}
\subsection{Training Analysis}

\begin{figure*}[t] 
	\centering
	\begin{subfigure}[t]{.49\textwidth} 
		\centering
		\includegraphics[width=\textwidth]{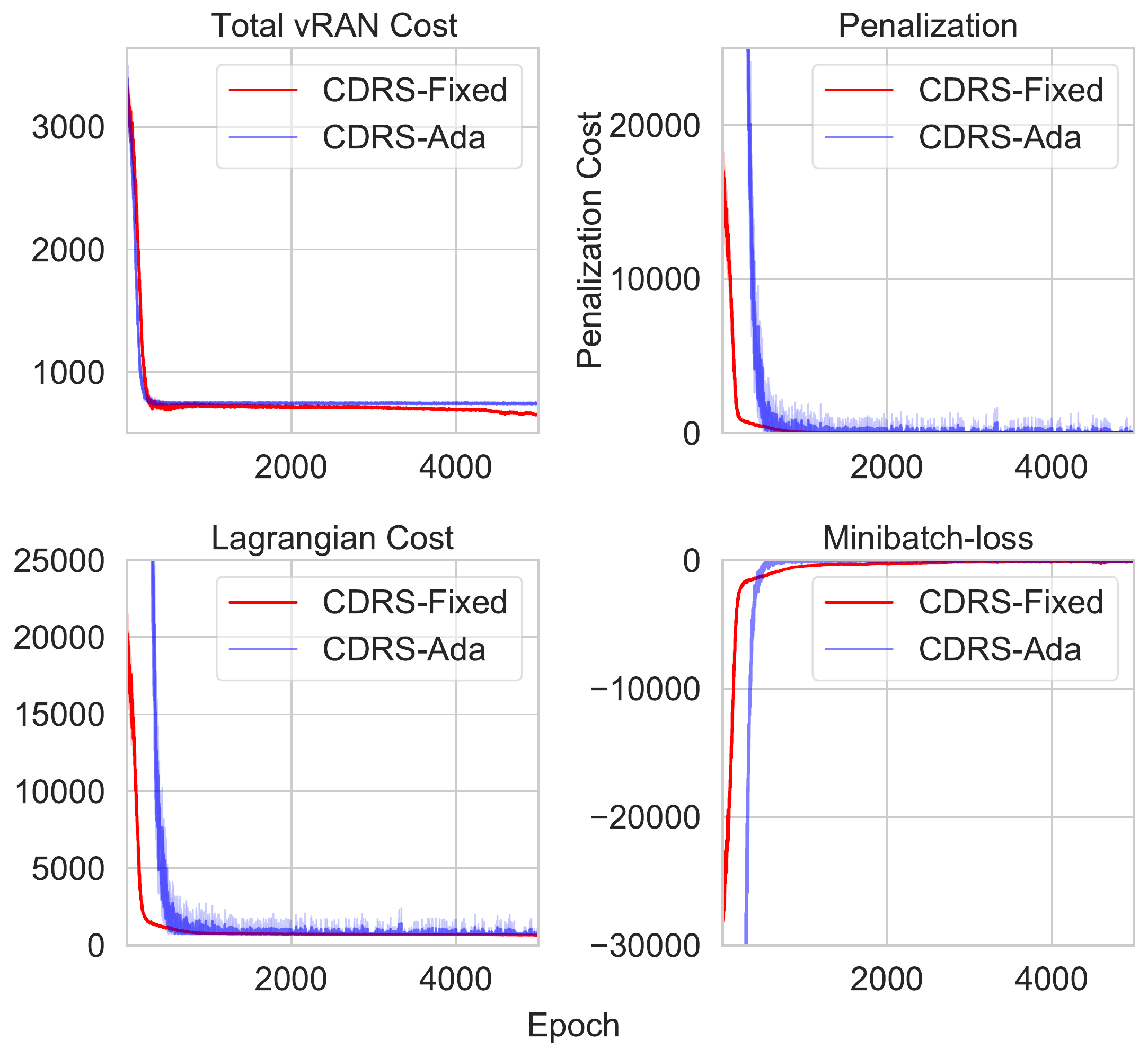}
		\small\caption{\small R1}
	\end{subfigure}
	\begin{subfigure}[t]{.49\textwidth} 
		\centering
		\includegraphics[width=\textwidth]{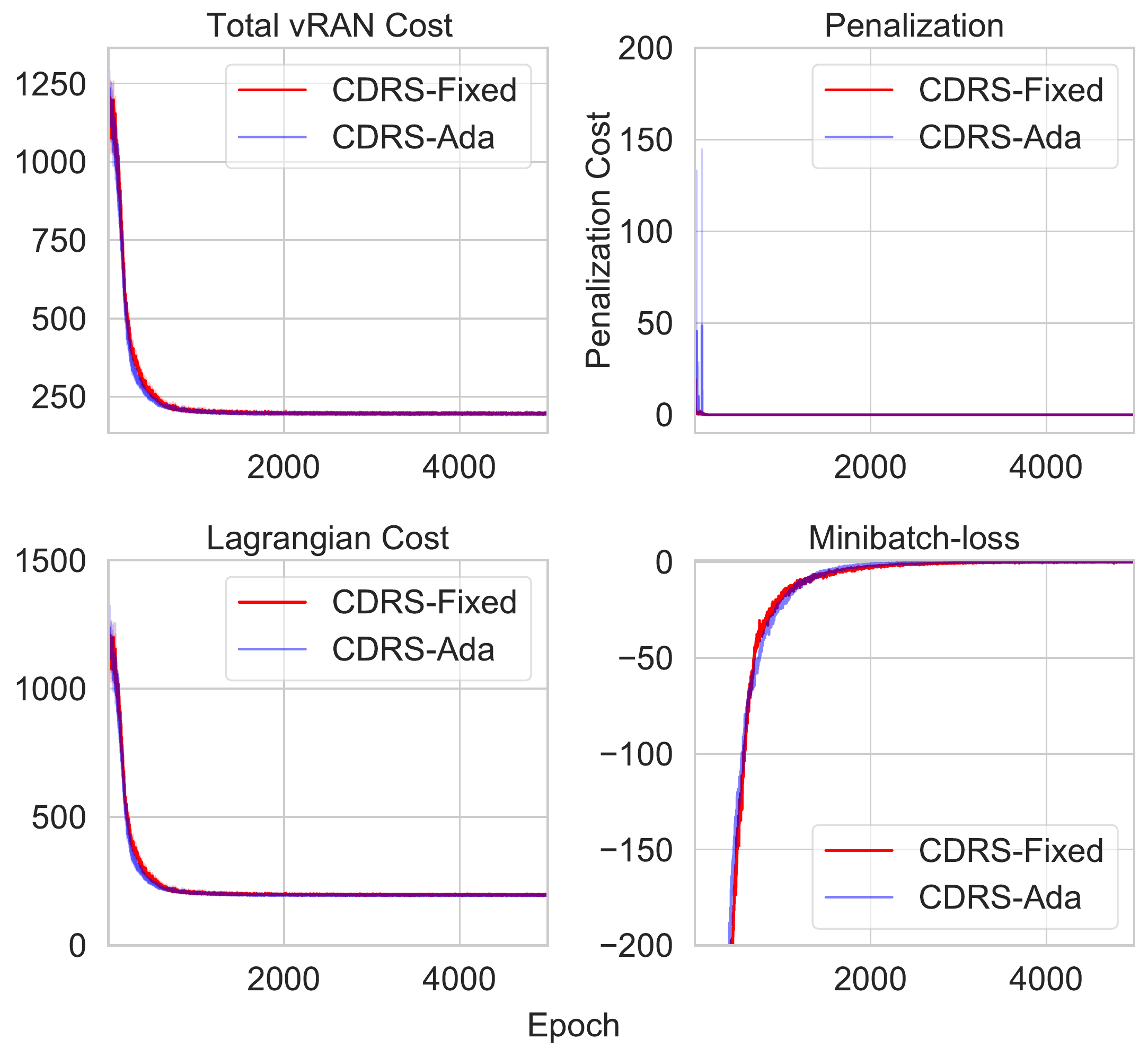}
		\small\caption{\small R2}
	\end{subfigure}		
	\caption{\small \textbf{Training results of CDRS in (a) R1 and (b) R2.} CDRS-Fixed uses a fixed value of penalty coefficient (reward shaping) with $\mu_i = 1, \forall i$. CDRS-Ada utilizes an adaptive update of penalty coefficient.} 
	\label{fig:res_train} 
	\vspace{-3mm}
\end{figure*}

We aim to examine the behaviour of CDRS-Fixed and CDRS-Ada during the training process in R1 and R2. We focus on the mini-batch loss, reward (total network cost), Lagrangian cost and penalization.  

Fig. \ref{fig:res_train} visualizes the training of CDRS-Fixed and CDRS-Ada in R1 and R2. We found additional costs because of penalization at the beginning of the training for both settings in R1 and R2. It occurs because CDRS-Fixed and CDRS-Ada try to find the solution, but violate the constraint sets (e.g., latency, bandwidth, computation). Fig. \ref{fig:res_train} also shows a significant difference in the cost of penalization in R1 compared to R2. The main reason is that R1 has stricter constraints, e.g., larger path delays, smaller link capacity than R2. We can also see that CDRS-Fixed and CDRS-Ada improve their policy by focusing on constraint satisfaction and then correcting the weights via stochastic gradient descent. It is proven from our agent's behaviour in R1 and R2, where each penalization cost keeps decreasing and turns to zero as soon as the training goes. CDRS-Ada sets the penalty coefficient increasing in the ascent direction, causing a higher penalization value than CDRS-Fixed. However, it can help speed up the policy toward constraint satisfaction, i.e., CDRS-Ada penalization downs faster than CDRS-Fixed.

We also found that the policy of CDRS-Ada converges faster than CDRS-Fixed from the behaviour of mini-batch loss in R1. Despite the mini-batch loss decreases to near zero after several epochs, the mini-batch loss of CDRS-Ada diminishes faster than CDRS-Fixed. However, CDRS-Ada suffers from more severe sub-optimality. It is shown by the total vRAN cost of CDRS-Ada that converges to a fixed value but has a higher cost compared to CDRS-Fixed.  Then, we have the Lagrangian cost from the sum of vRAN cost and penalization cost. It describes how our agent tries to minimize the primal problem $\mathbb{P}_{\text{1P}}$ through the dual problem $\mathbb{P}_{\text{1D}}$.  When our agent finally dismisses the penalization cost, it means that all constraints are satisfied. As a result, the Lagrangian cost becomes equal to the vRAN cost, and the penalty coefficient of CDRS-Ada converges to a fixed value.  Although having different behaviours, CDRS-Ada and CDRS-Fixed can learn the solution and converge to the local minima or saddle point in R1 and R2.

\textbf{Findings:} 1) R1 has stricter constraint requirements than R2; hence, it produces a higher additional cost for penalization to CDRS-Fixed and CDRS-Ada. 2) CDRS-Fixed and CDRS-Ada improve the policy by focusing on the penalization; then, it adjusts the weights as the training goes. 3) CDRS-Ada receives higher penalization compared to CDRS-Fixed as a result of increasing the penalty coefficient in the ascent direction; however, it also helps speed up the policy to constraint satisfaction. 4) CDRS-Ada converges faster but has a higher cost than CDRS-Fixed in R1. 5) When all constraints are satisfied,  the Lagrangian cost becomes equal to the total vRAN cost, and the penalty coefficient of CDRS-Ada converges to a fixed value.

\vspace{-2mm}
\subsection{Accuracy of Solutions}
%
%
\begin{figure*}[t]
	\centering
	\begin{subfigure}[t]{.47\textwidth}
		\centering
		\includegraphics[width=\textwidth]{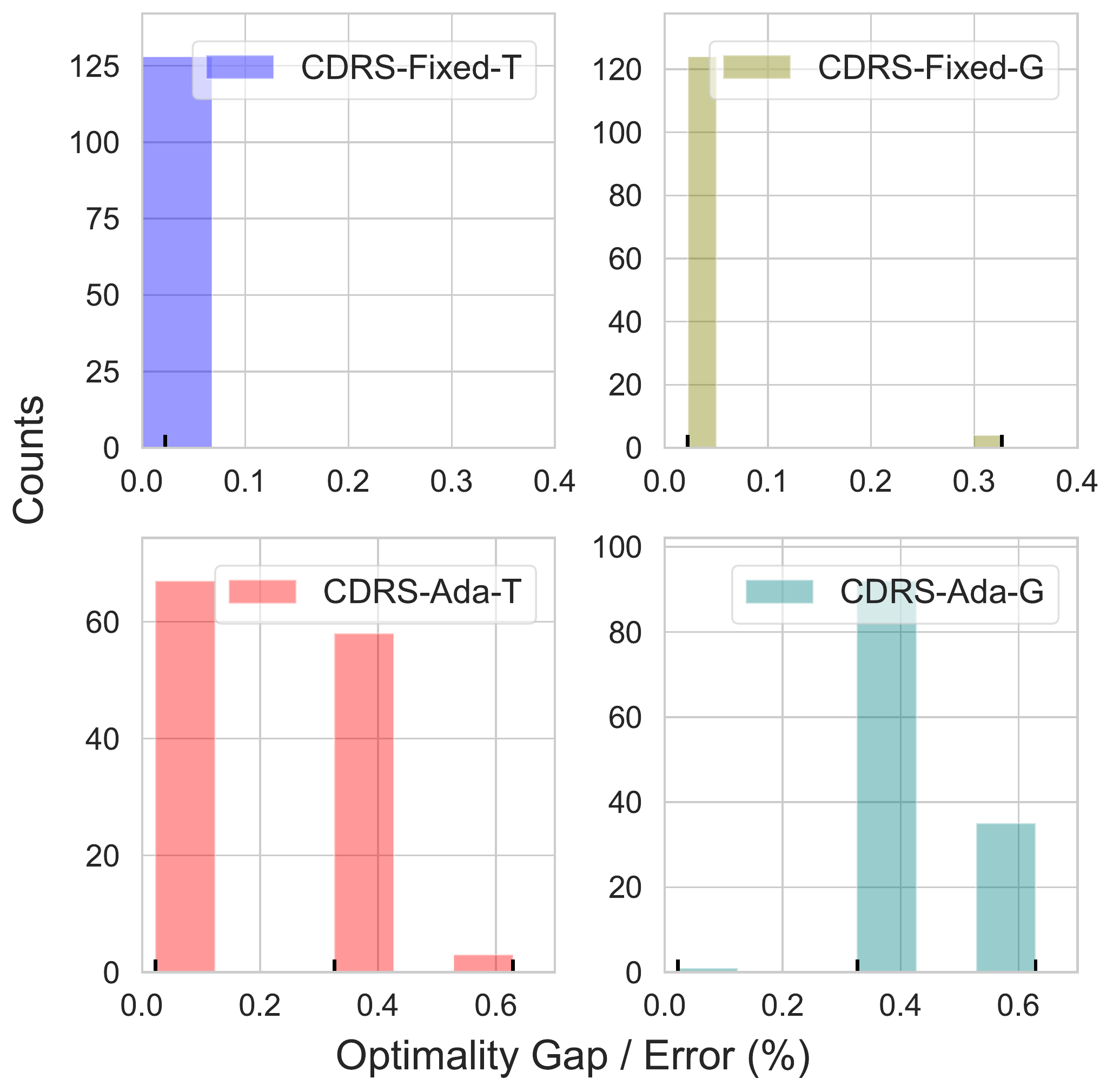}
		\small\caption{\small R1}
	\end{subfigure}
	\begin{subfigure}[t]{.47\textwidth}
		\centering
		\includegraphics[width=\textwidth]{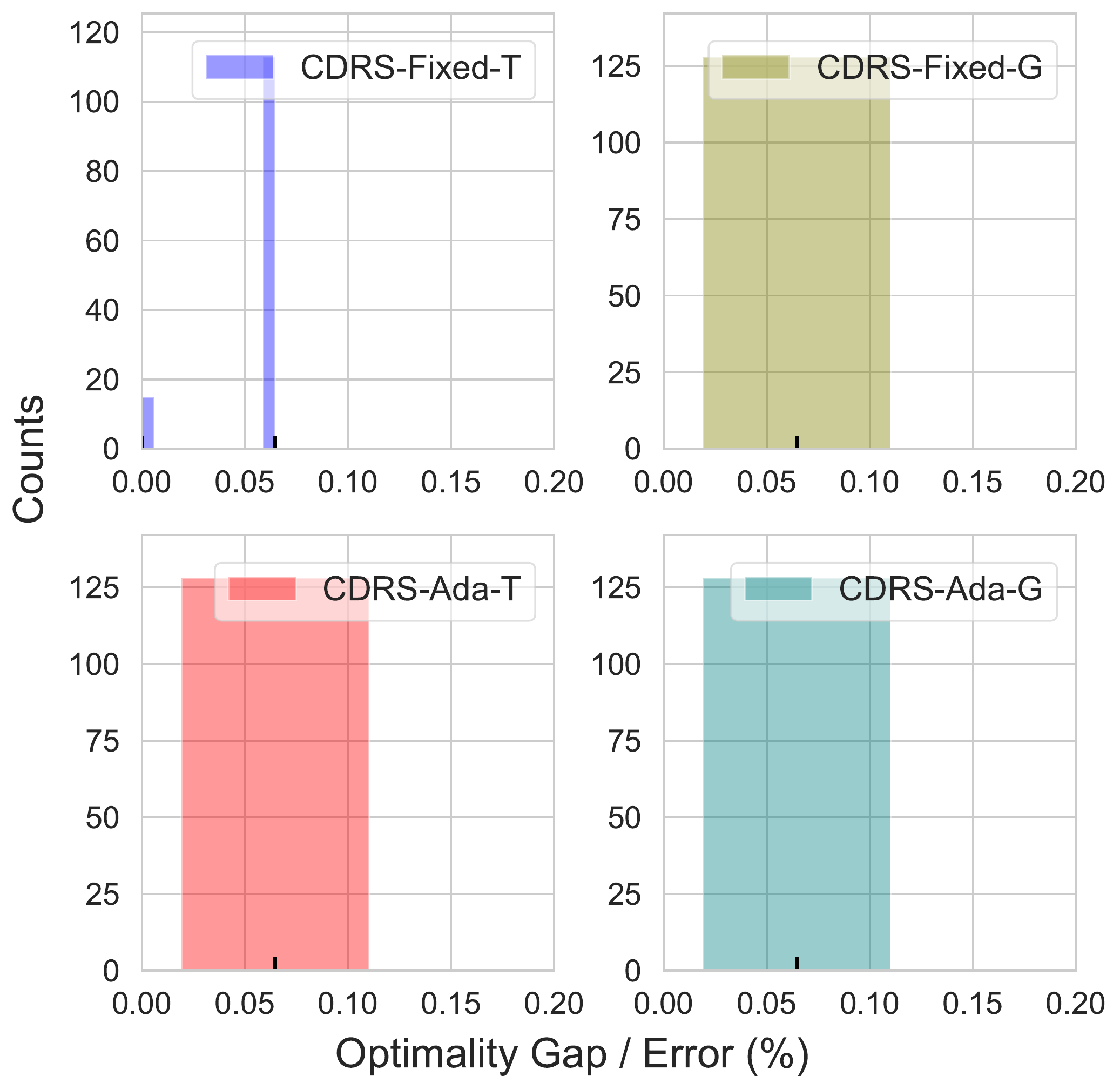}
		\small\caption{\small R2}
	\end{subfigure}	
	\caption{\small \textbf{Histogram of CDRS accuracy in (a) R1 and (b) R2.} The accuracy is calculated over 128 tests. CDRS-Ada-T and CDRS-Fixed-T are set with $T=15$ and $16$ samples.} 	\label{fig:_accmain}
	\vspace{-3mm}	
\end{figure*}

In this part, we study the accuracy of CDRS over different penalty coefficient and search strategy settings: CDRS-Fixed-G, CDRS-Fixed-T, CDRS-Ada-G and CDRS-Ada-T. We conduct 128 tests with a distinct sequence order of the BSs in R1 and R2 to assess how accurate these four CDRS settings find the solution of the vRAN split problem. We utilize three pretraining models \secrev{from} our CDRS training.

Fig. \ref{fig:_accmain} shows the distribution of \secrev{the solutions from} CDRS-Fixed-G, CDRS-Fixed-T, CDRS-Ada-G and CDRS-Ada-T in R1 and R2. \secrev{Each bar counts the number of offered solutions resulting in some suboptimality, represented using the optimality gap (error). It shows that the distribution varies between four settings, especially in a stricter environment (R1). Still,} all of these settings can guarantee less than $0.6 \%$ (R1) and $0.1 \%$ (R2) of the optimality gap. In R1, CDRS-Fixed-G and CDRS-Fixed-T perform better by offering lower solution errors ($ \leq 0.05 \%$ and $\leq 0.05 \%$ of optimality gap) than CDRS-Ada-G and CDRS-Ada-T ($\leq 0.6 \%$). It means that a fixed penalty coefficient setting can lead to a better optimality performance during the test than the adaptive one. However, CDRS-Fixed-G, CDRS-Ada-G and CDRS-Ada-T have a similar performance in R2. Regardless of R1 or R2, using a sampling method with a temperature hyperparameter can improve (or at least at same) the optimality performance than greedy decoding. It is shown from the higher total number of solutions (counts) for a sampling method that having a lower error. The combination of a fixed penalty in the training and temperature sampling method (CDRS-Fixed-T) can improve the solution performance significantly both in R1 and R2. It can achieve an optimal value (R2) and less than $0.05 \%$ of error for a more complex environment (R1). It is also shown that CDRS-Fixed-T is less affected to the stricter environment than any other settings where all of the distribution solutions are in less than $0.05 \%$.

\textbf{Findings:} 1) CDRS-Fixed-G, CDRS-Fixed-T, CDRS-Ada-G and CDRS-Ada-T can guarantee the solution with very close to the optimal value offering less than $0.6 \%$ (R1) and $0.1 \%$ (R2) of the optimality gap over 128 tests. 2) CDRS-Fixed-T can significantly improve the optimality performance (offers $\leq 0.05 \%$ of optimality gap) and outperforms the other settings.

\vspace{-2mm}
\subsection{Impact of Routing Cost}
\vspace{-1mm}

This part studies the impact of altering the routing cost to CDRS-Fixed-G, CDRS-Fixed-T, CDRS-Ada-G and CDRS-Ada-T. We aim to examine how the routing cost affects optimality performance and the total network cost. Hence, the default routing cost is changed within a range of $\gamma=0.1$ to $\gamma=1$. This change can arise due to increasing or decreasing the leasing agreement's price, maintenance, etc. The traffic load is fixed with $\lambda_{n}=150$ Mbps. We utilize three \secrev{pretraining} models, conduct 128 tests for each routing cost scale, and analyze the offered solutions' distribution. We also consider benchmarking with two extremes of RAN setups, fully D-RAN and C-RAN\footnote{We practically can not implement C-RAN because our RANs do not meet the constraint requirements of delay, bandwidth and CU capacity to deploy C-RAN. The presented C-RAN in this experiment is just for benchmarking; hence we also do not consider the penalization cost (constrains violation) for this case.} to assess how significant the routing cost affects the total network cost over various RAN setups.

\begin{figure*}[t] 
	\centering
	\begin{subfigure}[t]{.99\textwidth}
		\centering
		\includegraphics[width=\textwidth]{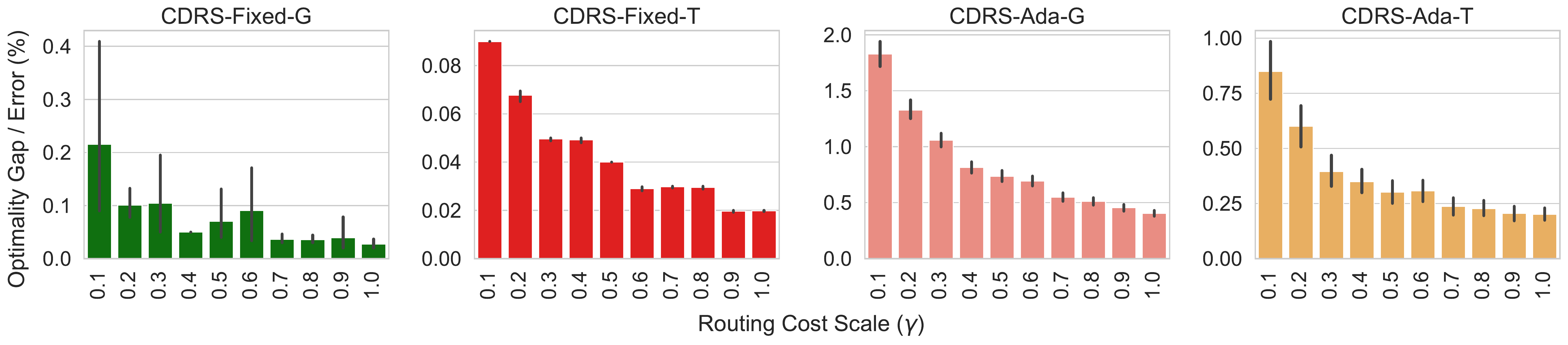}
		\small\caption{\small R1}
	\end{subfigure}
	\begin{subfigure}[t]{.99\textwidth}
		\centering
		\includegraphics[width=\textwidth]{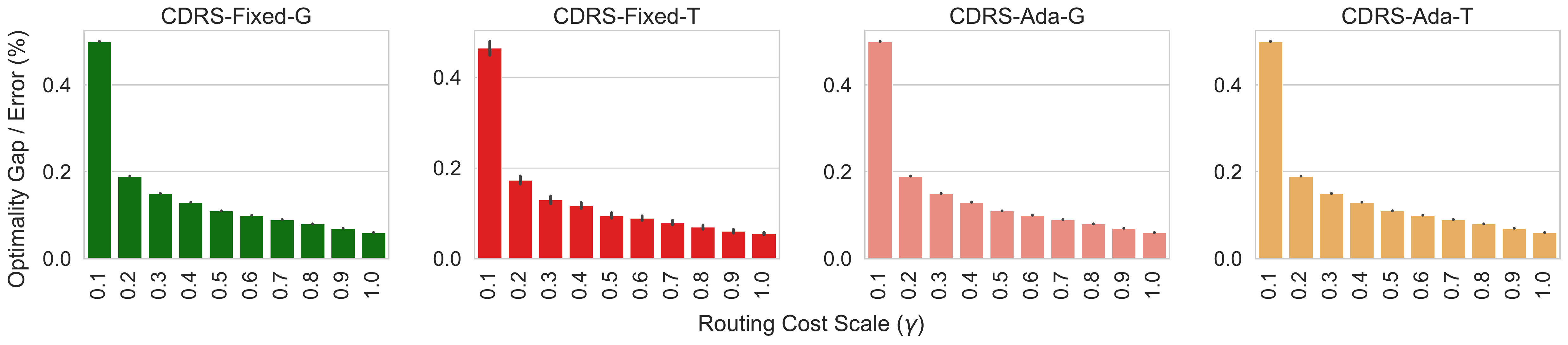}
		\small\caption{\small R2}
	\end{subfigure}		
	\caption{\small \textbf{Impact of the routing cost to the accuracy in (a) R1 and (b) R2.} Study of altering the routing cost to the optimality performance with $\lambda_{n}=150$ Mbps, $\forall n \in \mathcal{N}$. There are 128 tests for each routing scale $[0.1,1]$.} \label{fig:routing_acc}
	\vspace{-3mm}
\end{figure*}

\begin{figure*}[t] 
	\centering
	\begin{subfigure}[t]{.375\textwidth}
		\centering
		\includegraphics[width=\textwidth]{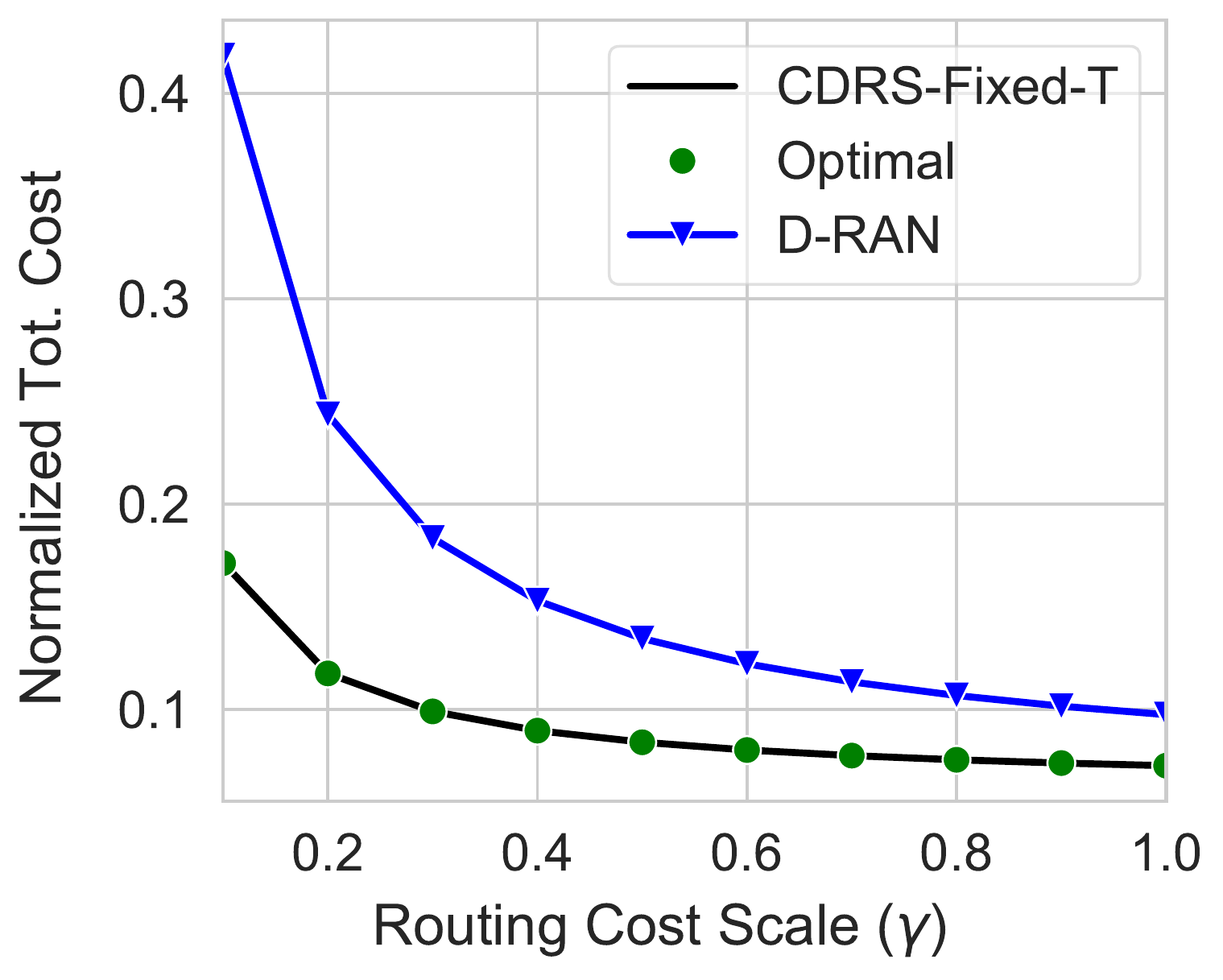}
		\small\caption{\small R1}
	\end{subfigure}
	\begin{subfigure}[t]{.375\textwidth}
		\centering
		\includegraphics[width=\textwidth]{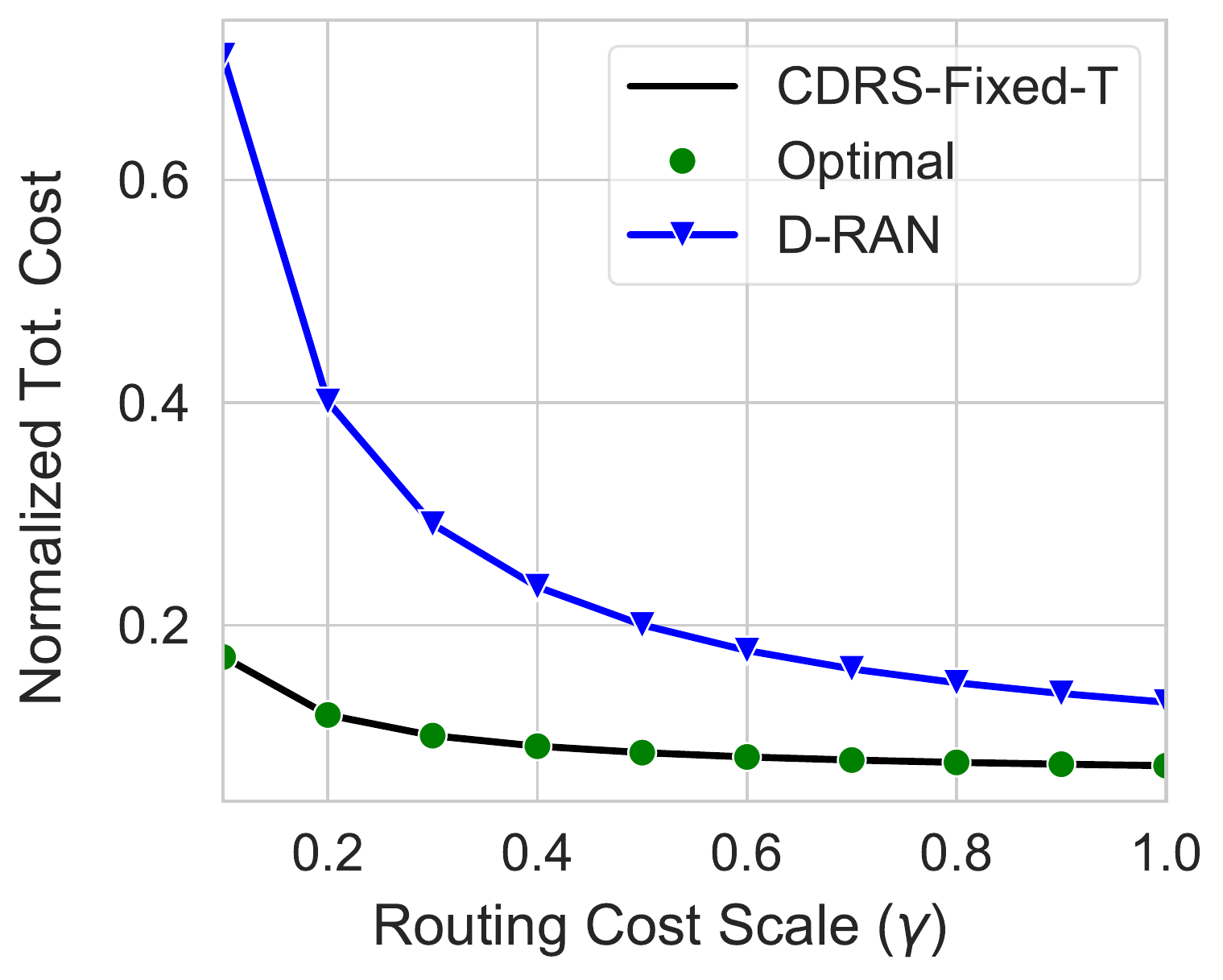}
		\small\caption{\small R2}
	\end{subfigure}		
	\caption{\small \textbf{Impact of routing cost to the total cost in (a) R1 and (b) R2.} We also compare our approach (e.g., CDRS-Fixed-T) to two extreme cases: fully D-RAN and C-RAN, and the optimal value with the routing cost scaling from 0.1 to 1 of default R1 and R2. The presented cost above is normalized toward fully C-RAN cost.} \label{fig:routing_cost}
	\vspace{-3mm}
\end{figure*}

Fig. \ref{fig:routing_acc} depicts how the routing cost affects the optimality performance of CDRS-Fixed-G, CDRS-Fixed-T, CDRS-Ada-G and CDRS-Ada-T. It shows that the \secrev{overall} optimality gap (error) diminishes as the routing cost increases; then, it converges to a specific value. In R1, we see a performance improvement as the errors decrease for CDRS-Fixed-G ($\approx75\%$), CDRS-Fixed-T ($\approx75\%$), CDRS-Ada-G ($\approx78\%$) and CDRS-Ada-T ($\approx75\%$) \secrev{by the increase of routing cost}. It also shows that CDRS-Ada-G gets the most impact while CDRS-Fixed-T is the least affected. In R2, all CDRS settings \secrev{also} have a similar trend in terms of error \secrev{performance}. Although we have changed the routing cost from the default parameter, we found that \secrev{altering the} routing cost gives relatively less effect to these settings where the errors are maintained under $1.8 \%$. CDRS-Fixed-T even can guarantee the solution under $0.08 \%$ ($\gamma  = 0.1$) of the optimality gap.

Fig. \ref{fig:routing_cost} shows the routing cost's effect on the total network cost of CDRS-Fixed-T and D-RAN, normalized to the C-RAN cost in R1 and R2. \secrev{It shows that CDRS-Fixed-T can obtain a larger cost-saving than the D-RAN cost at a cheaper routing cost by up to $59.06\%$ of cost-saving at $\gamma = 0.1$ while only $25.49\%$ of cost-saving at $\gamma = 1$ in R1. Compared to C-RAN, CDRS-Fixed-T can save the cost by up to $92\%$ at $\gamma = 1$ in R1. However, this gain diminishes as the routing cost decreases and eventually CDRS-Fixed-T will reach near the C-RAN cost if all constraint requirements are eligible. A similar trend also appears for R2. Moreover, CDRS-Fixed-T can offer the solution extremely close to the optimal solution by $\leq 0.09\%$ (R1) and $\leq 0.5\%$  (R2). }
%


\textbf{Findings:} 1) The increase of routing cost \secrev{reduces} the optimality gap \secrev{(error)}; then, \secrev{it} converges to a fixed value. 2) CDRS-Fixed-T is the least affected by the routing cost changes, while CDRS-Ada-G is the most affected. 3) \secrev{Scaling} the routing cost from $\gamma = 0.1$ to $\gamma = 1$ does not significantly degrade the optimality performance. 4) CDRS-Fixed-T has the lowest optimality gap  \secrev{than other CDRS settings}, and becomes the most cost-effective setup in R1 and R2. \secrev{5) CDRS-Fixed-T can reach near the D-RAN cost at a high routing cost, while it can be near the C-RAN cost at a cheap routing cost if all constraint requirements are eligible.}

\vspace{-2mm}
\subsection{Impact of Traffic Load}
\vspace{-1mm}

\begin{figure*}[t] 
	\centering
	\begin{subfigure}[t]{.99\textwidth}
		\centering
		\includegraphics[width=\textwidth]{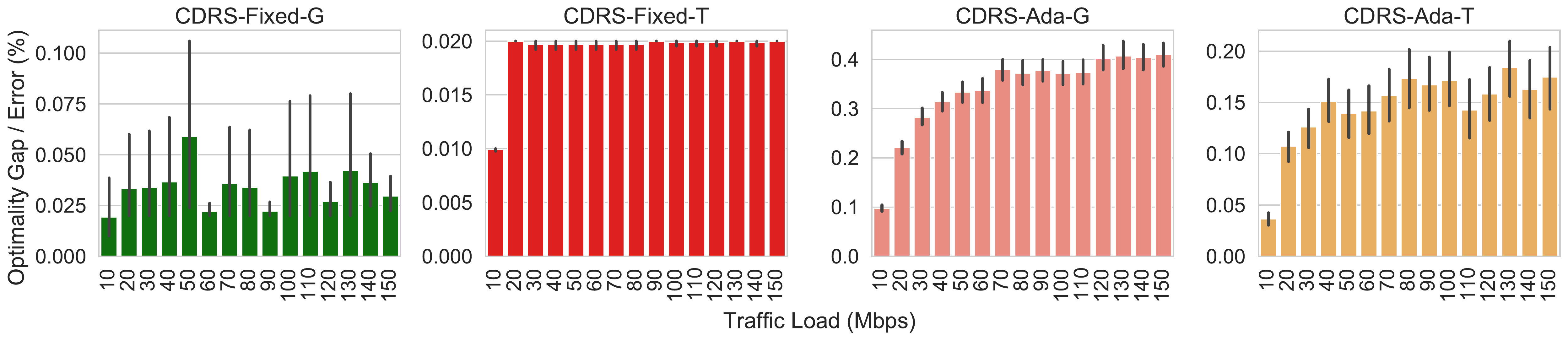}
		\small\caption{\small R1}
	\end{subfigure}
	\begin{subfigure}[t]{.99\textwidth}
		\centering
		\includegraphics[width=\textwidth]{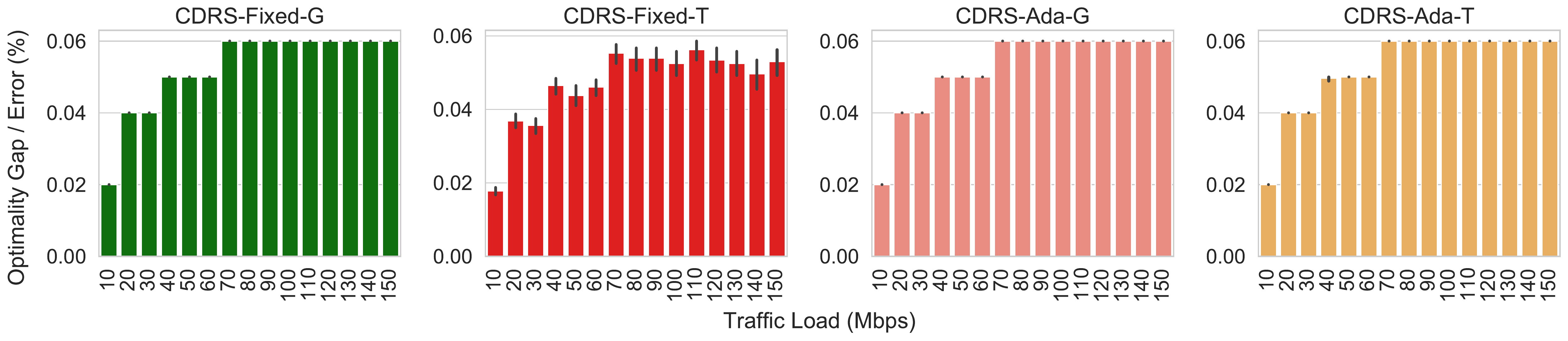}
		\small\caption{\small R2}
	\end{subfigure}		
	\caption{\small \textbf{Impact of the traffic load to the accuracy in (a) R1 and (b) R2.} Study of traffic load to the optimality performance. There are 128 tests for each traffic load. } \label{fig:traffic_acc}
	\vspace{-3mm}
\end{figure*}

\begin{figure*}[t] 
	\centering
	\begin{subfigure}[t]{.375\textwidth}
		\centering
		\includegraphics[width=\textwidth]{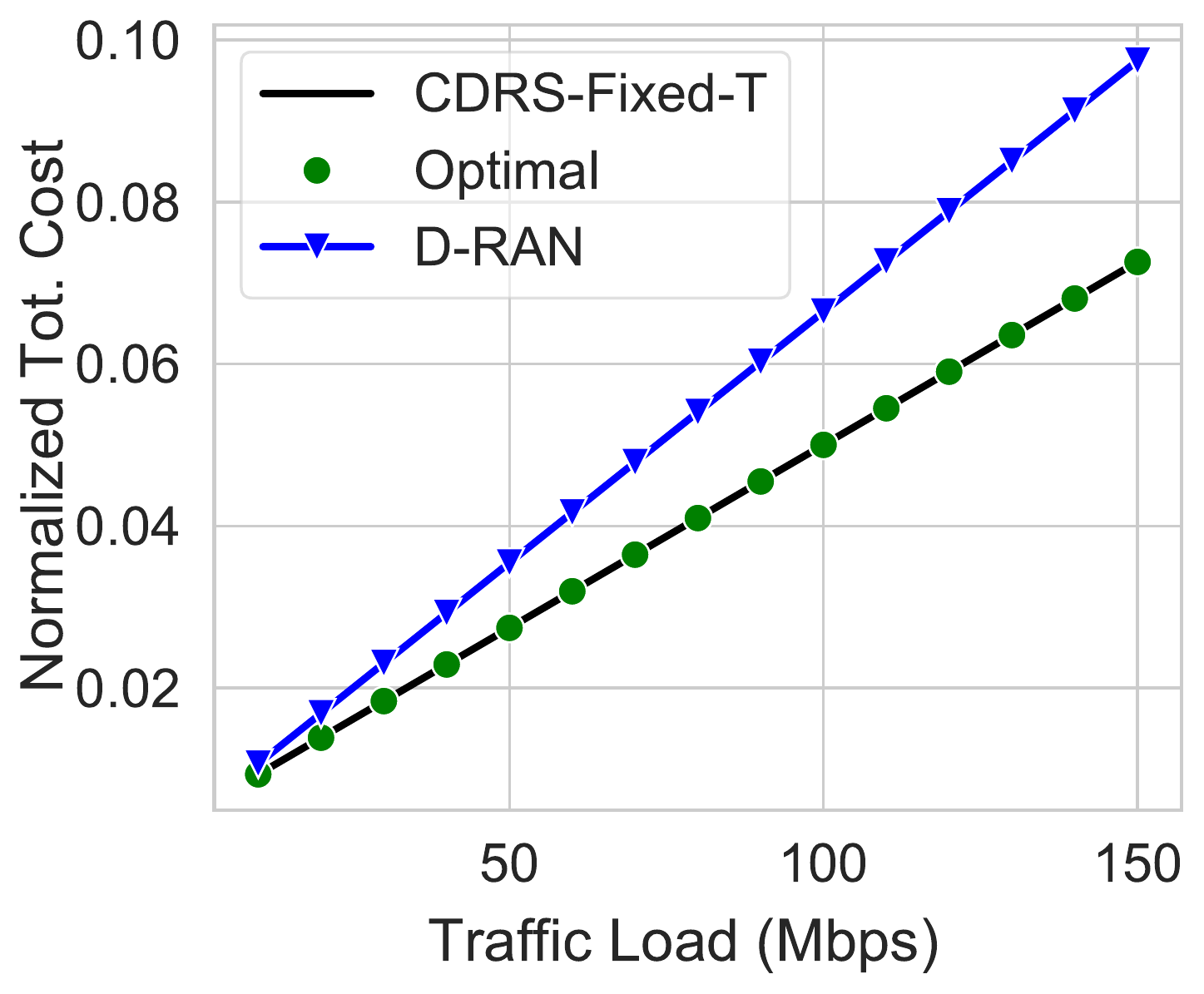}
		\small\caption{\small R1}
	\end{subfigure}
	\begin{subfigure}[t]{.375\textwidth}
		\centering
		\includegraphics[width=\textwidth]{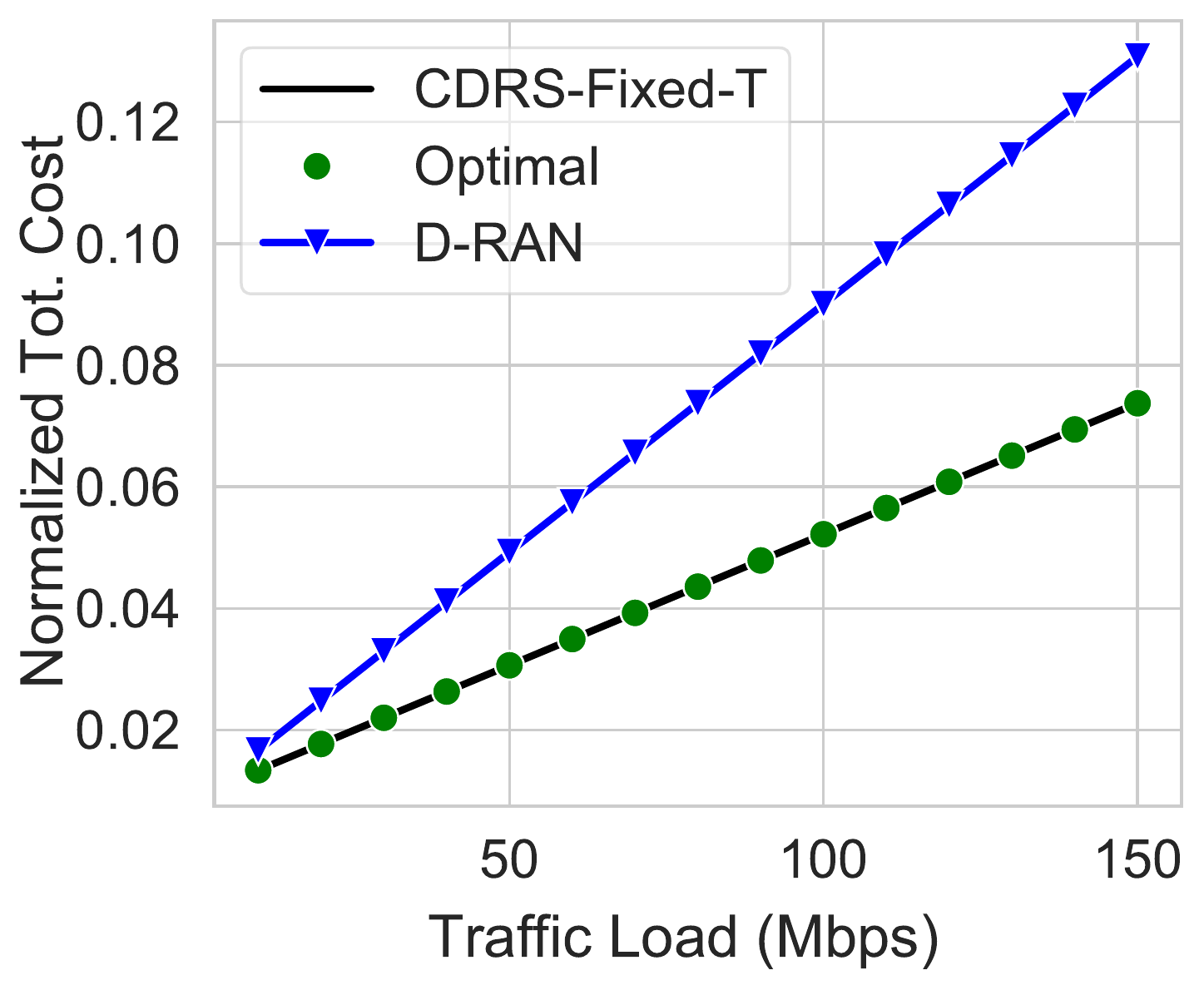}
		\small\caption{\small R2}
	\end{subfigure}		
	\caption{\small \textbf{Impact of traffic load to total vRAN cost in (a) R1 and (b) R2.} On the comparison of our approach (e.g., CDRS-Fixed-T) to fully D-RAN. The presented cost above is normalized toward fully C-RAN cost.} \label{fig:traffic_cost}
	\vspace{-3mm}
\end{figure*}

In this part, we assess how the traffic load affects the optimality performance and the total network cost. We change the traffic load from 10 Mbps to 150 Mbps. This evaluation is conducted using three \secrev{pretraining} models and examined over 128 tests.   

Fig \ref{fig:traffic_acc} shows the impact of altering the traffic load to the optimality performance of CDRS-Fixed-G, CDRS-Fixed-T, CDRS-Ada-G and CDRS-Ada-T. In R1, it shows that the increase of traffic load in line with the rise of the error to CDRS-Ada-G and CDRS-Ada-T, but it then diminishes to a fixed value, i.e., around $0.4 \%$ (CDRS-Ada-G) and $0.18\%$ (CDRS-Ada-T). However, the traffic load does not significantly affect CDRS-Fixed-G and CDRS-Fixed-T, where they stay at around $0.04\%$ and $0.02\%$ of errors, respectively, in R1. In R2, CDRS-Fixed-G, CDRS-Fixed-T, CDRS-Ada-G and CDRS-Ada-T have the same trend where the optimality gap increases with the traffic load; then, it diminishes at around  $0.05\%$. We also found that CDRS-Fixed-T \secrev{has a} better optimality performance and a more stable solution.

Fig \ref{fig:traffic_cost} examines the impact of traffic load on CDRS-Fixed-T and D-RAN cost normalized to the C-RAN cost. Despite an increase in CDRS-Fixed-T cost as the traffic load rises, it shows that CDRS-Fixed-T is still the most cost-effective compared to D-RAN and C-RAN \secrev{in} R1 and R2. \secrev{CDRS-Fixed-T almost has the same cost as D-RAN at the low traffic load with only $12.33\%$ cost-saving. This cost-saving then increases for the higher traffic load settings by up to $25.5\%$ at 150 Mbps in R1. This trend also happens in R2. Compared to C-RAN, CDRS-Fixed-T significantly outperforms at the low traffic load, but this gain then diminishes as the increase of the load. CDRS-Fixed-T can reach near the C-RAN cost when all constraint requirements are satisfied, and the traffic load is high, but the routing cost is significantly low.}

%

\textbf{Findings:} 1) CDRS-Fixed-T can offer to better optimality performance and more stable solution \secrev{than other CDRS settings}. 2) In R2, all CDRS settings have similar trends where the increase of traffic load can also increase the optimality gap, but it then diminishes and stays at around $0.05\%$ for CDRS-Fixed-T and $0.06\%$ for the others. 3)  CDRS-Fixed-T is the most cost-efficient compared to C-RAN and D-RAN. 4) \secrev{CDRS-Fixed-T can eventually almost have the same C-RAN cost when all constraint requirements are satisfied, and the traffic load is high, but the routing cost is significantly low.}

\vspace{-2mm}
\subsection{Computational Time}
\vspace{-1mm}

Finally, we examine the computational time to solve a single instance of the vRAN split problem. We use a small laptop with an Intel Core i5-7300U CPU@2.60GHz and 8GB memory. The computational time for each CDRS setting is a result of averaging 128 executions with a trained model. We report this evaluation in Table \ref{table:computationaltime}.  Overall, our proposed CDRS settings: CDRS-Fixed-G, CDRS-Fixed-T, CDRS-Ada-G and CDRS-Ada-T, have a faster computational time than the MIP solver. CDRS-Ada-G is the fastest with $0.0120$ secs and $0.0077$ secs in R1 and R2 reaching to $22.82$ times faster than the MIP solver. We also found that any CDRS settings with greedy decoding for the inference process, e.g., CDRS-Fixed-G, CDRS-Ada-G, is more time-efficient than a temperature sampling method with around 10-20 times faster. It is also shown that CDRS-Ada-G/T has a slightly faster computational time than CDRS-Fixed-G/T. Finally, we can sort from the fastest computational time as 1) CDRS-Ada-G, 2) CDRS-Fixed-G, 3) CDRS-Ada-T, 4) CDRS-Fixed-T, 5) the MIP solver.

\begin{table*}[t!] \centering
	\begin{small}
		\begin{tabular}{@{}cccccc@{}}\toprule
			\textbf{Topology}& \textbf{MIP solver} &\textbf{CDRS-Fixed-T} & \textbf{CDRS-Fixed-G}  & \textbf{CDRS-Ada-T} & \textbf{CDRS-Ada-G}
			\\ \midrule
			\textbf{R1} &      0.2527   & 0.2026 & 0.0155& 0.1985 & 0.0120          
			\\ \hdashline
			{\textbf{R2}} &  0.1756 &  0.1240 & 0.0098 & 0.1207 &0.0077
			\\ \hdashline
			\bottomrule
		\end{tabular}
	\end{small}
	\caption{\small \textbf{Computational time.} Study of computational time for solving a single problem instance in seconds. The presented computational time is a result of averaging 128 executions.}
	\label{table:computationaltime}
	\vspace{-3mm}
\end{table*}

\textbf{Findings:} 1) CDRS-Ada-G, CDRS-Fixed-G, CDRS-Ada-T, and CDRS-Fixed-T can reach up to $22.82, 17.99, 1.45$ and $1.41$ times faster than the MIP solver. 2) Greedy decoding is more time-efficient than a temperature sampling method for the inference process.

\vspace{-2mm}
\section{Conclusion} \label{sec:conclusion}
\vspace{-1mm}
In this paper, we have investigated the functional split optimization problem i\thirdrev{n which the BS functions can be deployed at the CU or DUs.} We have formulated the problem mathematically and analyzed the complexity, which is shown to be combinatorial and NP-hard. 
Because finding the exact solution is computationally expensive and \thirdrev{precise modelling the actual vRAN system is highly non-trivial}, we have proposed CDRS \thirdrev{as a solution framework to optimize the functional splits of the BSs amidst minimal assumptions about the underlying system}.
\thirdrev{We have developed CDRS using} a chain rule-based stochastic policy to handle the interdependence between split decisions and the large action space. \thirdrev{We have applied} LSTM networks-based sequence-to-sequence model to approximate the policy. \thirdrev{Since this policy is limited to an unconstrained problem, and vRAN's constraint requirements bound each function placement decision}, we have leveraged a constrained policy gradient method to train the policy. We have \thirdrev{also provided} a search strategy \thirdrev{by greedy decoding or temperature sampling to improve the optimality performance at the test time}. 
\thirdrev{The performance of CDRS has been} extensively evaluated using synthetic and real network datasets. The results have shown that CDRS successfully learns the functional split decision with less than 0.05\% optimality gap, \thirdrev{attains considerable cost savings} compared to \thirdrev{C-RAN or D-RAN systems}, and has a faster computational time than the optimal baseline. 


\bibliographystyle{IEEEtran}
\bibliography{IEEEabrv,ref-vran-01}

\end{document}